\documentclass[a4paper,11pt]{article}

\usepackage[height=8.85in,width=6.75in]{geometry}

\usepackage{graphicx,rotating}
\usepackage[bookmarksopen,colorlinks=true,linkcolor=light_blue,citecolor=light_pink,urlcolor=dark_red,linktoc=all]{hyperref}
\usepackage{amsmath,mathtools,amsfonts,amssymb,slashed,braket,bm,bbm,cite,xcolor,bbold,physics}
\usepackage[footnotesize]{caption}

\usepackage{fourier}

\makeatletter
\@addtoreset{equation}{section}

\makeatother

\newcommand{\Mpl}{M_{\text{Pl}}}

\usepackage{multicol}
\definecolor{dark_red}{rgb}{0.7, 0., 0.}
\definecolor{light_pink}{rgb}{1,0.4,0.4}
\definecolor{light_blue}{rgb}{0.284602,0.317763,0.963947}
\definecolor{darkgreen}{RGB}{0, 100, 0}
\definecolor{desy_blue}{HTML}{009EE2}
\definecolor{desy_orange}{HTML}{FD8800}
\definecolor{forestgreen}{HTML}{228B22}
\definecolor{ochre}{HTML}{CCAA2B}

\begin{document}
\hypersetup{pageanchor=false}
\begin{titlepage}

\begin{center}

\hfill KEK-TH-2458\\
\hfill KEK-Cosmo-0298\\
\hfill KEK-QUP-2022-0020\\
\hfill TU-1171\\
\vskip 0.5in
{\Huge \bfseries  Formation of hot spots around\\ 
small primordial black holes\\}
\vskip .8in
{\Large Minxi He$^a$, Kazunori Kohri$^{a,b,c,d}$, Kyohei Mukaida$^{a,b}$, Masaki Yamada$^{e,f}$}

\vskip .3in
\begin{tabular}{ll}
$^a$& \!\!\!\!\!\emph{Theory Center, IPNS, KEK, 1-1 Oho, Tsukuba, Ibaraki 305-0801, Japan}\\
$^b$& \!\!\!\!\!\emph{Graduate University for Advanced Studies (Sokendai), }\\[-.3em]
& \!\!\!\!\!\emph{1-1 Oho, Tsukuba, Ibaraki 305-0801, Japan}\\
$^c$& \!\!\!\!\!\emph{QUP, KEK, 1-1 Oho, Tsukuba, Ibaraki 305-0801, Japan}\\
$^d$& \!\!\!\!\!\emph{Kavli IPMU (WPI), University of Tokyo,
Kashiwa, Chiba 277-8568, Japan}\\
$^e$& \!\!\!\!\!\emph{Frontier Research Institute for Interdisciplinary Sciences, Tohoku University, }\\[-.3em]
& \!\!\!\!\!\emph{6-3 Azaaoba, Aramaki, Aoba-ku, Sendai 980-8578, Japan }\\
$^f$& \!\!\!\!\!\emph{Department of Physics, Tohoku University, }\\[-.3em]
& \!\!\!\!\!\emph{6-3 Azaaoba, Aramaki, Aoba-ku, Sendai 980-8578, Japan}\\
\end{tabular}

\end{center}

\vskip .4in

\begin{abstract}
\noindent
\end{abstract}
In this paper, we investigate the thermalization of Hawking radiation from primordial black holes (PBHs) in the early Universe, taking into account the interference effect on thermalization of high energy particles, known as Landau-Pomeranchuk-Migdal (LPM) effect. Small PBHs with masses $ \lesssim 10^9 \, \mathrm{g} $ completely evaporate before the big bang nucleosynthesis (BBN). The Hawking radiation emitted from these PBHs heats up the ambient plasma with temperature lower than the Hawking temperature, which results in a non-trivial temperature profile around the PBHs, namely a hot spot surrounding a PBH with a broken power-law tail. We find that the hot spot has a core with a radius much larger than the black hole horizon and its highest temperature is independent of the initial mass of the PBH such as $2 \times 10^{9} \, {\rm GeV} \times (\alpha/0.1)^{19/3}$, where $\alpha$ generically represents the fine-structure constants. We also briefly discuss the implications of the existence of the hot spot for phenomenology. 
\end{titlepage}

\tableofcontents
\thispagestyle{empty}
\renewcommand{\thepage}{\arabic{page}}
\renewcommand{\thefootnote}{$\natural$\arabic{footnote}}
\setcounter{footnote}{0}
\newpage
\hypersetup{pageanchor=true}

\section{Introduction}
\label{sec:intro}
Hawking radiation~\cite{Hawking:1974rv,Hawking:1975vcx} is a direct consequence of quantum properties of fields around a black hole, through which the mass of the black hole is continuously carried away by the emitted particles that have a thermal spectrum characterized by the Hawking temperature. The evaporation becomes significant when getting close to the end of the process because both the Hawking temperature and the mass loss rate increase as the black hole mass decreases. The Hawking temperature can become as high as Planck scale at the last stage when the black hole mass is of order of Planck mass, which is of great interest for the strong relevance with physics at extremely high energy scale. 

However, black holes with astrophysical origins have masses larger than the solar mass, which indicates negligible effects of Hawking radiation within the age of the Universe. 
Primordial black holes (PBHs)~\cite{Zeldovich:1967lct,Hawking:1971ei,Carr:1974nx,Carr:1975qj} which form via the collapse of the overdense regions in the early Universe, on the other hand, theoretically can have masses ranging from Planck mass to that as large as supermassive black holes at galaxy centers, depending on the time of formation.%
\footnote{
A scenario of PBH formation from late collapse of false vacuum bubbles generated during inflation 
is considered in Refs.~\cite{Garriga:2015fdk,Deng:2017uwc}, in which cold and hot spots around a PBH are formed at a later epoch by a shock wave~\cite{Deng:2018cxb}. 
}
The allowed abundances of PBHs in different mass ranges are constrained by different observations (see the review~\cite{Carr:2020gox} and the references therein for more detail). 
Large PBHs ($ \gg 10^{15} \, \mathrm{g} $) are not much affected by Hawking radiation so that they can survive until today. They are considered as a good candidate of a considerable fraction of dark matter (DM)~\cite{Chapline:1975ojl} or of black holes detected by LIGO/Virgo Collaboration~\cite{LIGOScientific:2016aoc,Sasaki:2016jop}, which has been intensively studied recently. 

Small PBHs ($ \ll 10^{15} \,\mathrm{g} $), on the contrary, evaporate by the present time, but they may play an important role in the early Universe. 
The evaporating PBHs can directly produce all the particles within and beyond the Standard Model (SM), for example, gravitons~\cite{Hooper:2020evu} and supersymmetric particles, even when their masses are much larger than the temperature of the Universe.
It is also shown that the Hawking radiation can be responsible for the generation of baryon asymmetry~\cite{Carr:1976zz,Barrow:1981zv,Barrow:1990he,Baumann:2007yr,Fujita:2014hha,Hook:2014mla,Hamada:2016jnq} and have impacts on the axion dark matter models~\cite{Bernal:2021yyb,Bernal:2021bbv,Mazde:2022sdx}. 
A small PBH might induce the Higgs vacuum decay as well~\cite{Gregory:2013hja,Burda:2015isa,Burda:2015yfa,Burda:2016mou,Tetradis:2016vqb,Canko:2017ebb,Gregory:2020cvy,Hayashi:2020ocn}, where its physical interpretation and the validity of the calculation are still under debate~\cite{Gorbunov:2017fhq,Mukaida:2017bgd,Kohri:2017ybt,Shkerin:2021zbf,Shkerin:2021rhy,Strumia:2022jil}.  
Since the evaporation process accelerates as the black hole mass gets smaller, the completion of reheating by Hawking radiation can be ``sudden'' in a presumed PBH-dominated epoch, which can have non-trivial contribution to the gravitational wave background by enhancing the so-called induced gravitational wave~\cite{Inomata:2019ivs,Inomata:2020lmk}. 
The high temperature Hawking radiation may heat up the surrounding plasma and restore the symmetry at high energy scales which leads to the production of topological defects like monopoles~\cite{Das:2021wei}, and may trigger the electroweak sphaleron to affect the baryogenesis process.

In this paper, we discuss in depth the thermalization process of the high-energy particles from Hawking radiation in the plasma surrounding a PBH, which is essential for further discussion about the production of topological defects, vacuum decay, baryogenesis, and particle production by the heated-up plasma. 
We extend the analysis given by Ref.~\cite{Das:2021wei}, where they discussed the diffusion around a PBH but assuming an instantaneous thermalization for high-energy particles emitted through Hawking radiation.
Since the energy of emitted particles increases with time, it takes more and more time to get thermalized as the PBH mass becomes smaller and smaller.
The main purpose of this paper is to discuss this finite-time effect.
Specifically, we take into account the Landau-Pomeranchuk-Migdal (LPM) effect~\cite{Landau:1953um,Migdal:1956tc,Gyulassy:1993hr,Arnold:2001ba,Arnold:2001ms,Arnold:2002ja,Besak:2010fb} which is the bottleneck process for thermalization of high-energy particles in a low-temperature plasma (see Sec.\ref{sec:LPM}). 
The hard primaries from PBHs travel through the low-temperature plasma and emit soft population nearly collinearly to lose their energy through multiple scatterings. However, the emission is suppressed by the quantum interference of these scatterings, \textit{i.e.}, LPM effect. The high-energy mother particle needs long enough time to split into two energetic daughter particles to significantly reduce its energy. After a cascade of such kind of splittings, 
the LPM effect becomes irrelevant so these daughter particles get thermalized quickly and deposit their energy into the ambient plasma (see more explanation in, \textit{e.g.}, Refs.~\cite{Kurkela:2014tla,Harigaya:2013vwa,Mukaida:2015ria}).
Subsequently, the diffusion process tries to spread the energy into the ambient plasma. The competition among the time scales of evaporation, LPM effect, and diffusion results in a hot spot around an evaporating PBH with a broken power law tail, as we will show in Fig.~\ref{fig:Tprofile}. 
Because of a finite time scale of thermalization, 
the maximal temperature of the hot spot is finite and is 
of the order of $2 \times 10^{9} \, {\rm GeV} \times (\alpha/0.1)^{19/3}$, 
where $\alpha$ generically represents the fine-structure constants of SM interactions. 
This is significantly lower than the Planck scale even though the Hawking temperature is as large as the Planck scale just before the completion of evaporation. One has to be therefore careful about the meaning of temperatures for the Hawking radiation and the hot spot. The former represents the spectrum of Hawking radiation, which has nothing to do with the thermal equilibrium. The latter is the local temperature of thermal plasma, which is actually in equilibrium by efficient interactions. 
This hot spot may lead to interesting implications on phenomenology in cosmology. 

This paper is organized as follows. In Sec.~\ref{sec:basics_pbh}, we briefly review the basics of PBH evaporation and its cosmological evolution in the early Universe for the convenience of later discussion. We investigate the thermalization process of the Hawking radiation in Sec.~\ref{sec:therm_pbh}, specifying the essential physical effects during the whole process, \textit{i.e.}, the LPM effect and the thermal diffusion. We comment on possible implications of our results in Sec.~\ref{sec:implication} which can be the future directions of study. Finally, 
we summarize 
in Sec.~\ref{sec:conclusion}. 

\section{Basics of primordial black holes in the early Universe}
\label{sec:basics_pbh}

\subsection{Primordial black hole evaporation}
\label{sec:pbh_evap}
A black hole placed in an empty space loses its energy via thermal radiation, known as Hawking radiation.
The mass loss rate of an evaporating black hole with mass $M$ is given by
\begin{equation}
    \frac{\dd M}{\dd t} = - \frac{\pi \mathcal{G} g_{H\ast} \qty(T_\text{H})}{480} \frac{\Mpl^4}{M^2},
\end{equation}
where the reduced Planck mass is $\Mpl$,
the gray-body factor is denoted as $\mathcal{G} \simeq 3.8$,
the effective number of degrees of freedom of the Hawking radiation is $g_{H\ast}$,
and the Hawking temperature is 
\begin{equation}
    \label{eq:TBH}
    T_\text{H} = \frac{\Mpl^2}{M} \simeq 10^4 \,\mathrm{GeV} ~ \left( \frac{M}{10^9 \, \mathrm{g}} \right)^{-1}.
\end{equation}
A black hole whose Hawking temperature is greater than that of the ambient plasma continuously emits radiation and evaporates.
The evaporation time scale can be estimated as
\begin{equation}
    \label{eq:tev}
    t_\text{ev} (M) = \frac{160}{\pi \mathcal{G} g_{H\ast} \qty(T_\text{H})} \frac{M^3}{\Mpl^4} 
    \simeq 
    0.4 \,\mathrm{sec} ~ \left( \frac{g_{H\ast} (T_\text{H})}{108} \right)^{-1} \left( \frac{M}{10^9 \,\mathrm{g}} \right)^{3}.
\end{equation}
Assuming the radiation domination right after the PBH evaporation, one may readily translate it into the temperature of the Universe at the evaporation
\begin{equation}
    \label{eq:Tev}
    T_\text{ev} \simeq \left( \frac{\pi^2 g_\ast (T_\text{ev})}{90} \right)^{-\frac{1}{4}} 
    \left( \frac{\pi \mathcal{G} g_{H\ast} (T_\text{H}) }{320} \right)^{\frac{1}{2}}
    \left( \frac{\Mpl^5}{M_\text{ini}^3} \right)^{\frac{1}{2}}
    \simeq 1 \, \mathrm{MeV} ~ 
    \left( \frac{g_{\ast} (T_\text{ev})}{10.75} \right)^{-\frac{1}{4}}
    \left( \frac{g_{H\ast} (T_\text{H})}{108} \right)^\frac{1}{2} 
    \left( \frac{M_\text{ini}}{10^{9}\, \mathrm{g}} \right)^{-\frac{3}{2}},
\end{equation}
with $M_\text{ini}$ being the initial PBH mass, \textit{i.e.,} the PBH mass before the PBH evaporation becomes efficient.

In this paper, we consider PBHs that evaporate well before the big bang nucleosynthesis (BBN) to preserve the success of BBN, \textit{i.e.}, $T_\text{ev} \gtrsim 4$~MeV~\cite{Kawasaki:2000en,Hasegawa:2019jsa} or $M_\text{ini} \lesssim 10^9$~g (see Fig.18 of Ref.~\cite{Carr:2020gox}).
This implies
\begin{equation}
    t_\text{ev} \lesssim 1 \,\mathrm{sec}
    \quad \longleftrightarrow \quad   
    T_\text{H} \gtrsim 10 \, \mathrm{TeV}
    \quad \longleftrightarrow \quad
    T_\text{ev} \gtrsim 1 \,\mathrm{MeV}.
\end{equation}
In this case, the effective degrees of freedom is $g_{H\ast} (T_\text{H}) \simeq 108$ for the SM.
Hereafter, we take this as a fiducial value and often drop the dependence on $g_{H\ast}$ for brevity.

There are several mechanisms of PBH formation proposed so far.
As a representative example, let us consider the PBH formation in the radiation-dominated era.
Gravity can overcome the pressure if there exists an over-dense region at the horizon scale.
The typical masses of PBHs at formation can be expressed as a Hubble parameter at that time
\begin{equation}
    M_\text{ini} 
    = \gamma \frac{4 \pi \rho}{3 H_\text{ini}^3} 
    = 4 \pi \gamma \frac{\Mpl^2}{H_\text{ini}}
    \simeq 
    0.4 \, \mathrm{g} ~
    \left( \frac{\gamma}{0.2} \right) \left( \frac{H_\text{ini}}{6 \times 10^{13}\, \mathrm{GeV}} \right)^{-1}.
\end{equation}
Here, $H_\text{ini}$ is the Hubble parameter at the PBH formation, and $\gamma$ represents the fraction of the PBH mass in the horizon mass, which is estimated analytically for the PBH formation in the radiation-dominated era as $ \gamma \sim (1/\sqrt{3})^3 \simeq 0.2 $~\cite{Carr:1975qj}. 
The constraint on the tensor-to-scalar ratio by the CMB observation provides an upper bound on the Hubble parameter as $H \lesssim 6 \times 10^{13}\,\mathrm{GeV}$.
This in turn implies a lower bound on a PBH mass at its formation 
$M_\text{ini} \gtrsim 0.4 \, \mathrm{g}$.

In the following discussion of this paper, a hierarchy between the energy of emitted particles, $T_\text{H}$, and the temperature of the ambient plasma at evaporation, $T_\text{ev}$, plays a crucial role.
From Eqs.~\eqref{eq:TBH} and \eqref{eq:Tev}, one obtains
\begin{equation}
    \frac{T_\text{H}}{T_\text{ev}} 
    \geqslant 
    \frac{T_\text{H,ini}}{T_\text{ev}} 
    \sim 10^7 \times \left( \frac{M_\text{ini}}{10^9\,\mathrm{g}} \right)^\frac{1}{2}.
\end{equation}
For the initial PBH mass of our interest, \textit{i.e.,} $0.4 \, \mathrm{g} \lesssim M_\text{ini} \lesssim 10^9 \,\mathrm{g}$, we find a huge hierarchy between $T_\text{H}$ and $T_\text{ev}$. 
In particular, we always have $T_\text{H} > T_\text{ev}$.
The thermalization of such high-energy particles involves the LPM effect as we will discuss in Sec.~\ref{sec:therm_pbh}.

\subsection{Cosmological evolution}
\label{sec:evolution}
For the sake of completeness, let us briefly recap the cosmological evolution of PBHs after their formation.
Suppose that PBHs of mass $M_\text{ini}$ are formed in the radiation-dominated era with the energy fraction of
\begin{equation}
    \beta \equiv \frac{\rho_{\text{PBH,ini}}}{\rho_\text{tot,ini}} \simeq  \frac{\rho_{\text{PBH,ini}}}{\rho_\text{r,ini}}.
\end{equation}
The subsequent evolution depends on the value of $\beta$ as follows.
If the PBHs never dominate the Universe till their evaporation, the energy fraction at the evaporation can be expressed as
\begin{equation}
    \frac{\rho_\text{PBH,ev}}{\rho_\text{r,ev}} \simeq \beta \frac{T_\text{ini}}{T_\text{ev}},
\end{equation}
where $T_\text{ini}$ is the temperature of the Universe at the PBH formation.
On the other hand, this equality also indicates that the PBHs dominate the Universe for 
$\beta > T_\text{ev} / T_\text{ini} \sim 10^{-14} \left(M_\text{ini}/10^9 \, \mathrm{g}\right)^{-1}$.
In this case, the energy density of radiation produced from PBHs dominate over the primordial one, \textit{i.e.,} the dilute plasma is generated as
\begin{equation}
    \rho_\text{r} \sim \rho_\text{PBH} \frac{t}{t_\text{ev}}
    \quad \longrightarrow \quad
    \frac{\rho_\text{PBH,ev}}{\rho_\text{r,ev}} \simeq 1.
\end{equation}
Hence, one may conveniently summarize the energy fraction of PBHs at evaporation as follows
\begin{equation}
    \frac{\rho_\text{PBH,ev}}{\rho_\text{r,ev}} \simeq {\rm Min} \left[ 1, \beta \frac{T_\text{ini}}{T_\text{ev}} \right].
\end{equation}

For later usage, we estimate the mean separation of PBHs at  evaporation.
Since the PBHs behave as matter, we obtain
\begin{equation}
    \label{eq:distance}
    L_\text{PBH,ev} \sim \left( \frac{\rho_\text{PBH,ev}}{M_\text{ini}} \right)^{-\frac{1}{3}}
    \simeq 2\times 10^{-10} \, \mathrm{sec} ~  \left({\rm Min} \left[ 1, \beta \frac{T_\text{ini}}{T_\text{ev}} \right] \right)^{-\frac{1}{3}} 
    \left( \frac{g_{H\ast}}{108} \right)^{-\frac{2}{3}}
    \left( \frac{M_{\rm ini}}{10^9 \,\mathrm{g}} \right)^{\frac{7}{3}}.
\end{equation}

\section{Thermalization around a primordial black hole}
\label{sec:therm_pbh}

In this section, we discuss how high-energy particles from Hawking radiation become thermalized and dissipate into the ambient plasma. 
We determine the temperature profile as well as the maximal temperature of the hot spot around an evaporating PBH. 
For this purpose, in Sec.~\ref{sec:LPM} we briefly review the LPM effect, which is a bottleneck process for thermalization of high-energy particles. 
After they are thermalized, they tend to diffuse into the ambient plasma as we discuss in Sec.~\ref{sec:Diffusion}. 
The thermalization and the diffusion lengths for a corresponding time scale provide typical length scales for the temperature profile around a PBH, as we will see in Sec.~\ref{sec:therm_proc}. 

In the rest of this paper, we only provide order of magnitude estimations, \textit{i.e.}, $ \mathcal{O}(1) $ factors will be dropped and we also neglect all the logarithmic dependence. Those $ \mathcal{O}(1) $ numerical coefficients, for example in the LPM effect discussed below, depend on the details of all the particle contents in the plasma and also their interactions. 
Such model-dependent factors are not necessary for our qualitative understanding. 
More detailed treatment of these coefficients can be found in, \textit{e.g.}, Ref.~\cite{Arnold:2001ms}.

\subsection{Landau--Pomeranchuk--Migdal suppression}
\label{sec:LPM}

\begin{figure}[t]
	\centering
    \includegraphics[width=0.5\linewidth]{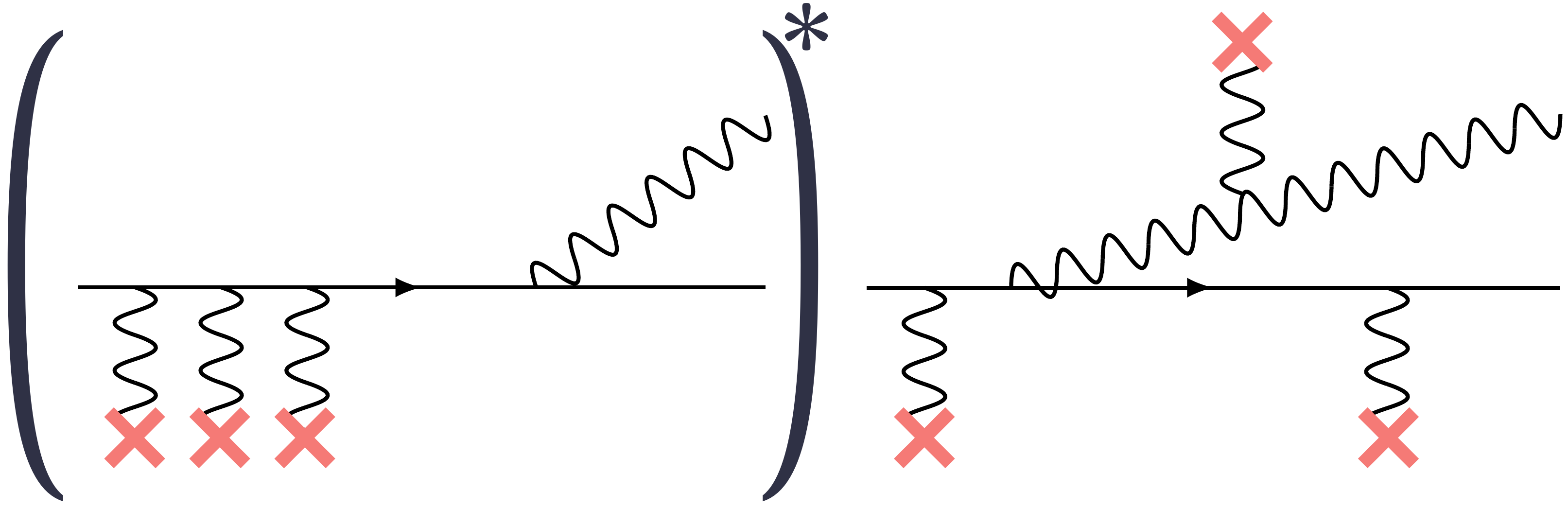} 
	\caption{
        Example of interference among a parent particle and daughter particles is shown. Such interferences lead to the LPM effect.
        Here the red cross represents the thermal plasma.
    }
	\label{fig:LPM}
\end{figure}

After high-energy particles with a momentum $ \sim T_\text{H}$ are emitted via the Hawking radiation, they deposit their energy into the ambient plasma of the low temperature $T_\text{ev}$.
Such energy-loss processes are dominated by nearly collinear splittings of a high-energy mother into low-energy daughters, where the daughters stay close to each other.
In this case, we cannot treat the subsequent scatterings incoherently, rather have to keep the quantum coherence until the overlap between the daughter particles is lost.
It is known that this quantum interference is destructive, which leads to the suppression of the splitting rate via the Landau--Pomeranchuk--Migdal (LPM) effect.
See also Fig.~\ref{fig:LPM}.

\begin{figure}[t]
	\centering
    \includegraphics[width=0.6\linewidth]{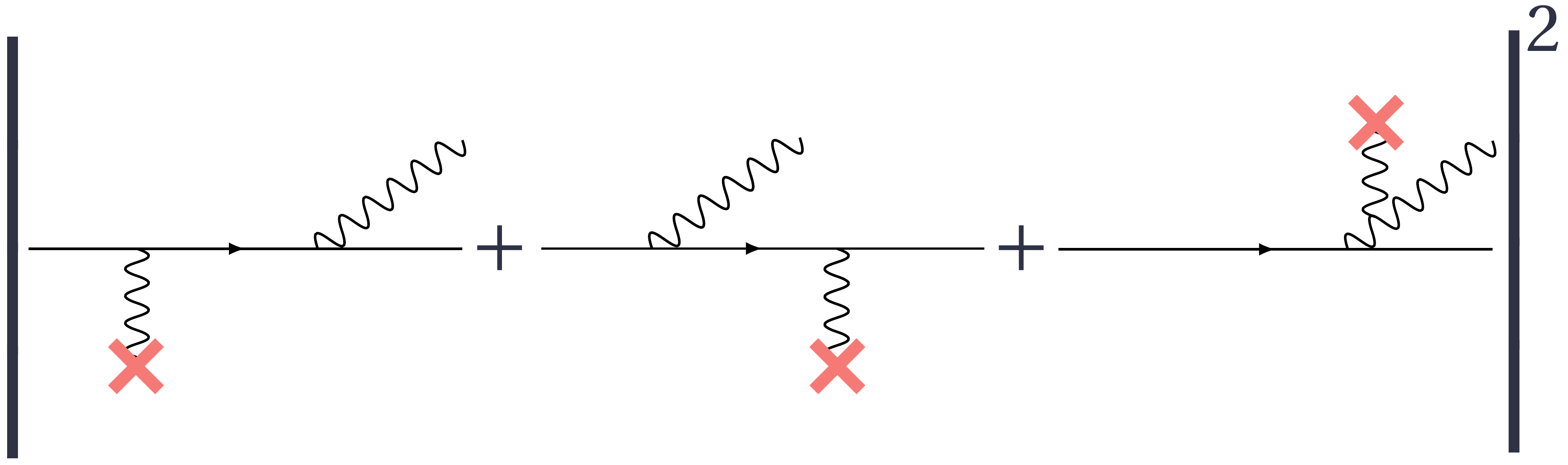} 
	\caption{
        Examples of diagrams responsible for inelastic scatterings are shown, which leads to Bremsstrahlung obtained by Bethe and Heitler.
        If the initial particle has a large energy, the interferences among these diagrams cancel the incoherent summation of the amplitude squared of each diagram.
    }
	\label{fig:BH}
\end{figure}

Here, instead of deriving the LPM-suppressed rate in SM (see \textit{e.g.,} \cite{Arnold:2002ja,Bodeker:2019ajh,Mukaida:2022bbo}), we provide an intuitive understanding of its typical parametric dependence following Ref.~\cite{Kurkela:2011ti} (see also Refs.~\cite{Harigaya:2014waa,Mukaida:2015ria,Mukaida:2022bbo}).
Suppose that an incoming high-energy particle with its energy $p$ scatters with the thermal plasma with the temperature $T$.
The large-angle elastic scatterings are suppresed by the large $p$ as $\alpha^2 T^3 / (p T) \sim \alpha^2 T^2 / p$.
On the other hand, the small-angle elastic scatterings are enhanced by the $t$-channel gauge-boson exchange, whose IR divergence is regularized by the medium induced mass of the gauge boson, \textit{i.e.}, $m_\text{th} \sim g T$.
Such small-angle elastic scattering rate can be estimated as $\alpha^2 T^3 / m_\text{th}^2 \sim \alpha T$.
It is tempting to consider an inelastic scattering as a small-angle elastic scattering followed by an emission of daughter particle (see Fig.~\ref{fig:BH}). Although this picture is not entirely correct because such small-angle scatterings lead to nearly collinear splittings with interferences, this would give the correct splitting rate when quantum coherence is negligible. The rate is originally obtained by Bethe and Heitler~\cite{Bethe:1934za},
\begin{equation}
    \Gamma_\text{BH} \sim \alpha \times \alpha T \sim \alpha^{2} T.
\end{equation}
Here, we collectively denote the relevant couplings as $\alpha$, for instance, the strong interactions, the top Yukawa interactions, the electroweak interactions, and so on.
Throughout this paper, we just take a common fiducial value of $\alpha \sim 0.1$ for simplicity, and hence we do not distinguish different $\alpha$'s.
However, this naive picture can be violated for nearly collinear splittings because of the destructive interference between the daughter particles (see Fig.~\ref{fig:LPM}).
In this case, the splittings does not take place unless the overlap between the daughter particles is lost.
For a daughter particle with a momentum $k$, the transverse spread of its wave is given by $1/k_\perp$, while its transverse velocity is $k_\perp / k$.
Hence, the time scale after which the overlap becomes negligible can be estimated as $t_\text{form} \sim k/k_\perp^2$. 
After $t \gtrsim t_\text{form}$, the high-energy mother particle can emit low-energy daughters with a probability of $\alpha$ which represents a coupling squared between the mother and the daughters.
Therefore, for $t_\text{form} \gtrsim 1/(\alpha T)$, 
the ordinary splitting after (or before) an elastic scattering with the thermal plasma (see Fig.~\ref{fig:BH}) is not possible.
This implies 
\begin{equation}
    \label{eq:split}
    \Gamma_\text{split} = {\rm Min} \left[ \Gamma_\text{LPM}, \Gamma_\text{BH} \right] , \qquad
    \Gamma_\text{LPM} \sim \alpha t_\text{form}^{-1} \sim \alpha \frac{k_\perp^2}{k},
\end{equation}
where $k$ is the energy of a daughter particle, and $k_\perp$ is the momentum of a daughter particle that is transverse to the incoming high-energy mother particle. 
The transverse momentum $k_\perp^2$ is determined by the diffusion of daughter particles transverse to the incoming high-energy mother particle, which is provided by the elastic scatterings mediated by the t-channel enhanced gauge bosons:
\begin{equation}
    k_\perp^2 \sim \hat q_\text{el} t_\text{form}, \quad \hat q_\text{el} \sim \alpha^2 \int_{\bm{p}} f(\bm{p}) \left[ 1 \pm f(\bm{p}) \right]
    \sim \alpha^2 T^3 ,
\end{equation}
where $ f(\bm{p}) $ is the distribution function of particles in the plasma which scatter with the high-energy mother particle. 
Together with $t_\text{form} \sim k / k_\perp^2$, we can write $t_\text{form}$ and $k_\perp^2$ in terms of $T$, $k$, and $\alpha$. 
Plugging them into Eq.~\eqref{eq:split}, we obtain the LPM-suppressed splitting rate:
\begin{equation}
    \Gamma_\text{LPM} \left( k,T \right) 
    \sim \alpha^2 T \sqrt{\frac{T}{k}}.
        \label{eq:LPM}
\end{equation}
We note that the splitting rate is smaller for a larger energy of daughter particle $k$
whereas the energy loss rate is larger for a larger $k$. 

Before investigating the detail of thermalization through LPM effect, we briefly explain the reason that the inelastic scattering with LPM suppression dominates over other processes during thermalization such as elastic scattering by comparing the rate of energy loss of the two processes. One can follow the analysis, for example, in Ref.~\cite{Harigaya:2014waa}. 
To achieve the thermalization, the incoming high-energy mother particle should lose its energy, whose time scale is characterized by the energy-loss rate.
The energy-loss rate of the LPM suppressed inelastic scattering can be estimated as 
\begin{equation}
    \frac{1}{E} \left. \frac{\dd E}{\dd t} \right|_{\rm inel} 
    \sim \frac{1}{E} \int^{E/2} \dd k\, \Gamma_\text{LPM} (k,T)
    \sim \alpha^2 T \sqrt{\frac{T}{E}} ~,
\end{equation}
where $E$ is the energy of the mother particle. 
As for the elastic scattering, the main contribution is the small-angle scattering mediated by the t-channel gauge boson exchange, whose rate is $\alpha T$.
Since the typical momentum transfer of the small-angle scattering is $m_\text{th} \sim g T$,
the high-energy mother particle loses its energy by $\sqrt{p/T} \times m_\text{th}^2 / \sqrt{p T} \sim \alpha T$, 
where a factor of $\sqrt{p/T}$ comes from the Lorentz boost factor for coordinate transformation from the center-of-mass frame to the lab frame. 
As a result, one can estimate the energy-loss rate as 
\begin{equation}
    \frac{1}{p}\left. \frac{\dd p}{\dd t} \right|_{\rm el} \sim \frac{\alpha T}{p} \times \alpha T \sim \alpha^2 T \sqrt{\frac{T}{p}} \times \sqrt{\frac{T}{p}} ~,
\end{equation}
which is suppressed by $ \sqrt{T/p} $ compared with the LPM case when $ p \gg T $. Therefore, LPM effect is essential in our consideration. 

Now we are ready to discuss the thermalization of high-energy mother particles with momenta $p$. 
The thermalization of a high-energy particle can be completed if it experiences multiple splittings into particles with energy of order the temperature. 
In terms of energy loss, the hard splitting into the largest $k$ dominates, \textit{i.e.,} $k \simeq p / 2$.
The LPM suppressed splitting rate \eqref{eq:LPM} is larger for a smaller $k$, so the daughter particles need less time for the subsequent splitting. 
Therefore the first splitting rate, $\Gamma_\text{LPM} (k \simeq p/2,T)$, determines the thermalization rate of high-energy particle. 
This motivates us to define the thermalization time of a high-energy mother particle of a momentum $p$ as 
\begin{equation}
    \label{eq:thermalization}
    t_\text{th} (p,T) \equiv \Gamma_\text{LPM} (p,T)^{-1}
    \qquad \text{for} \quad p \gtrsim T, 
\end{equation}
where we take $k \simeq p/2 \sim p$ in the argument for simplicity. 
For $t \gtrsim t_\text{th} (p, T)$, the highest-energy daughters with $k \simeq p /2$ break up and participate in the thermal plasma.
Note that, for $p \gtrsim T $, the thermalization is always LPM-suppressed.
As we have discussed at the end of Sec.~\ref{sec:pbh_evap}, this condition is always satisfied for $p = T_\text{H}$ and $T = T_\text{ev}$ in the parameter region of our interest.

To sum up, we have shown that the high-energy particles with momenta $T_\text{H}$ deposit their energy completely into the ambient plasma with temperature $T$ at $t_\text{th} (T_\text{H}, T)$.
Since they propagate with the speed of light for $t \lesssim t_\text{th}$, their energy is mostly released at the radius of $r \sim t_\text{th}$, which heats up the thermal plasma locally.
Still, the inner region of $r \lesssim t_\text{th}$ remains intact (see the left panel of Fig.~\ref{fig:schematic}).
Therefore, the thermal diffusion also plays an essential role for the global thermalization  (see also the middle panel of Fig.~\ref{fig:schematic}).

\subsection{Diffusion}
\label{sec:Diffusion}

\begin{figure}[t]
	\centering
 	\includegraphics[width=0.33\linewidth]{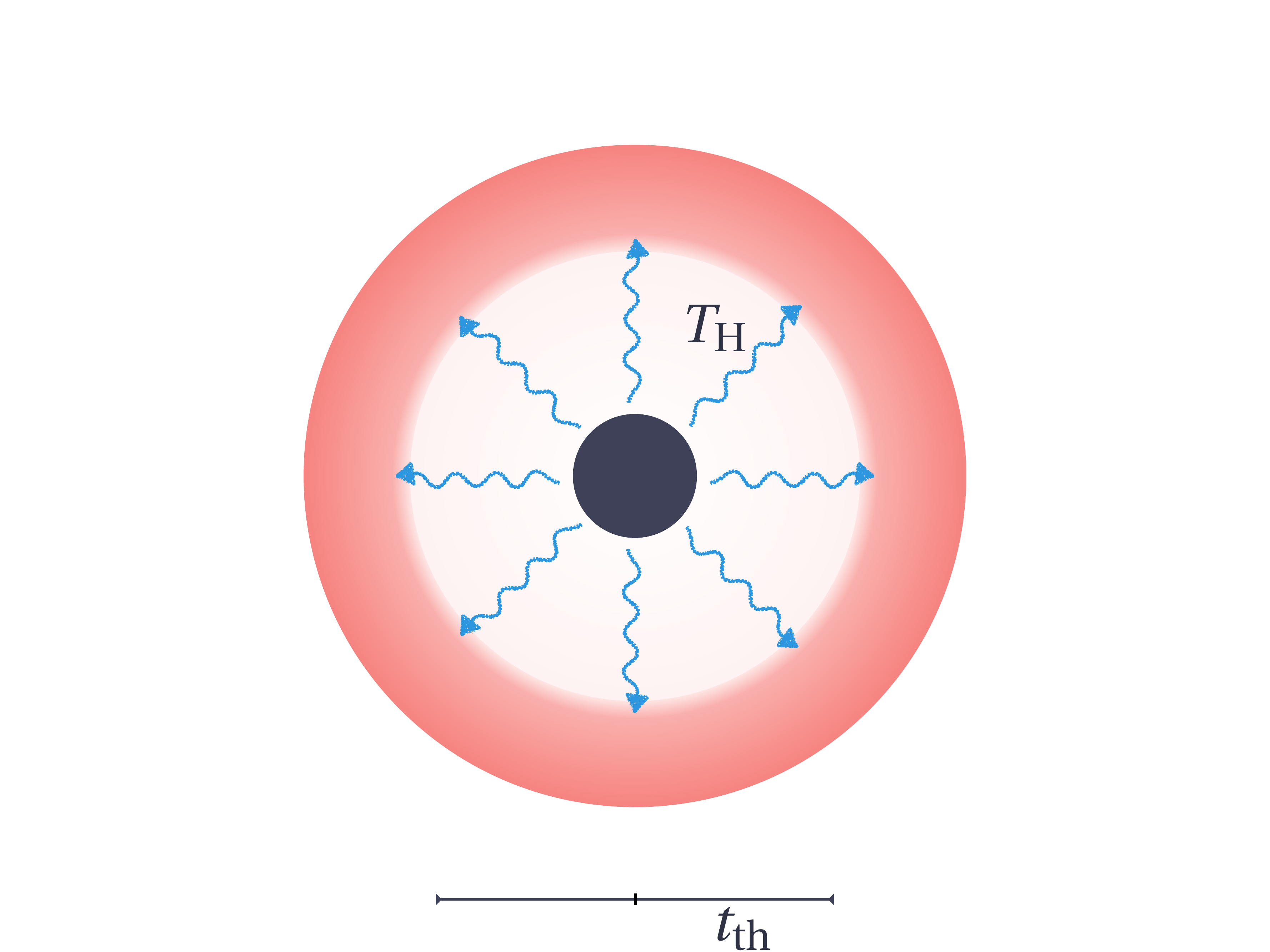}
    \includegraphics[width=0.33\linewidth]{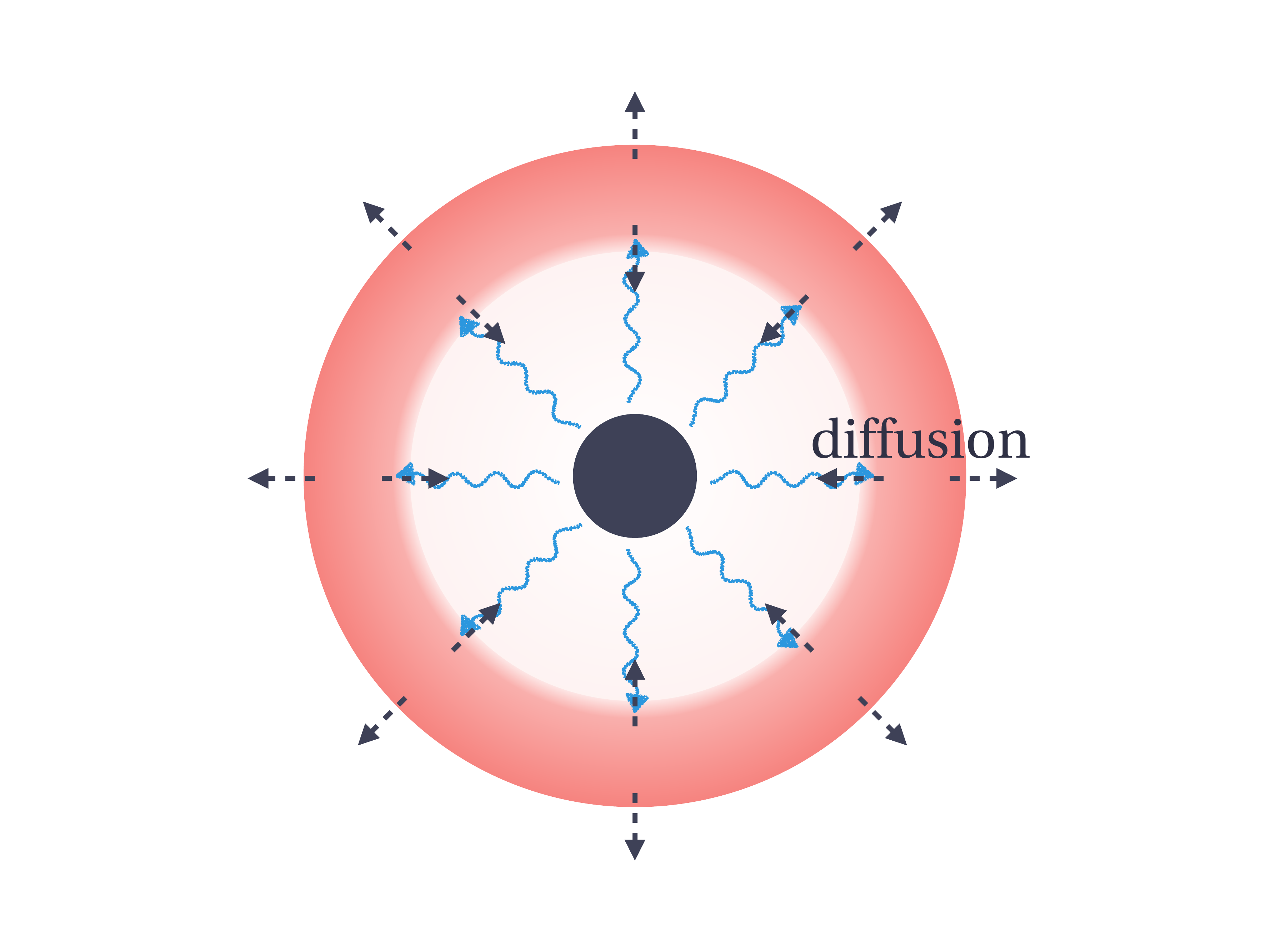} 
    \includegraphics[width=0.33\linewidth]{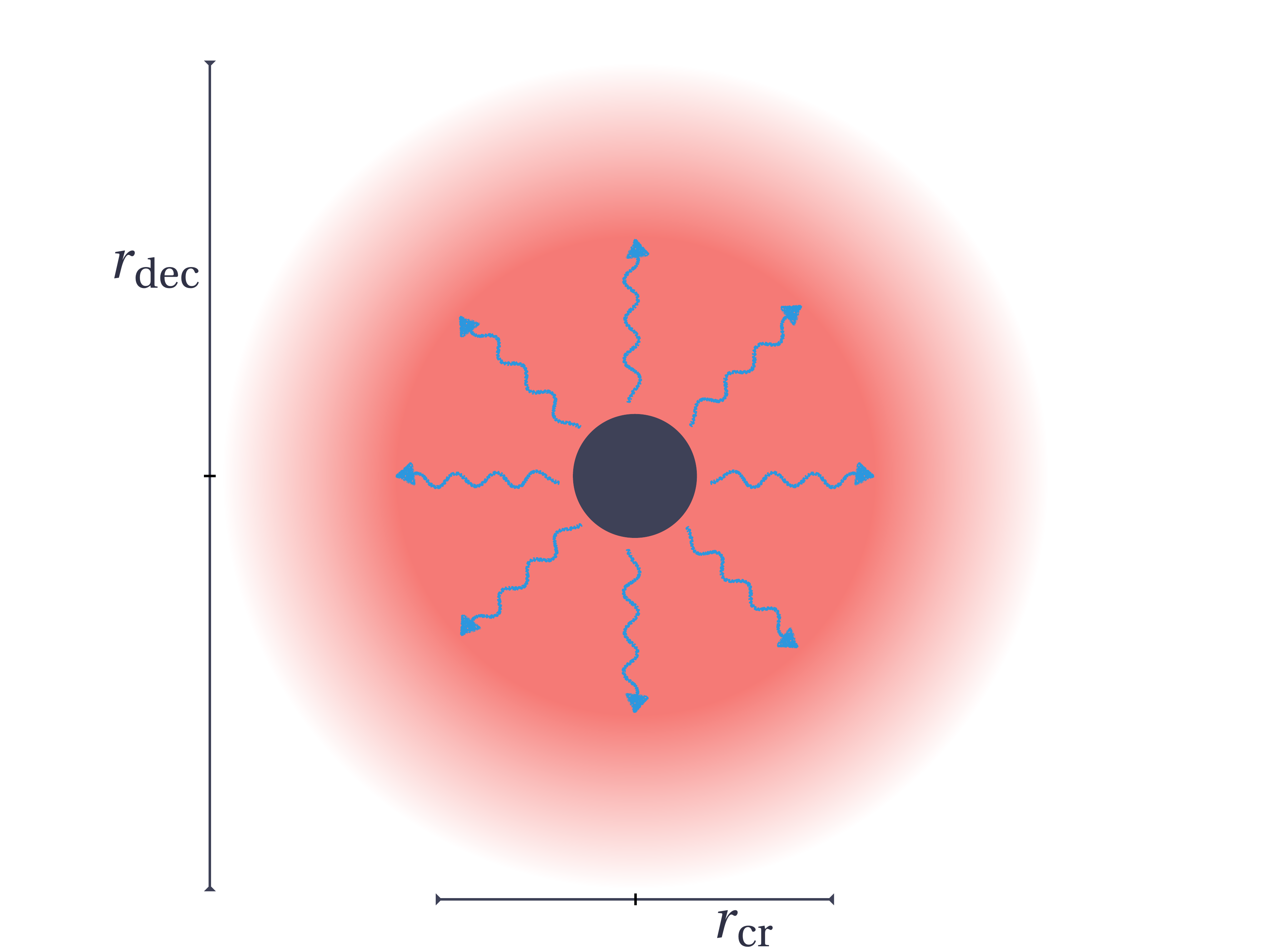} 
	\caption{
        Schematic picture for the formation of the hot spot is shown.
        (\textbf{Left panel}) The Hawking radiation with $T_\text{H}$ deposits their energy at $r \sim t_\text{th}$. 
        (\textbf{Middle panel}) The high-energy shell at $r \sim t_\text{th}$ spreads due to the diffusion.
        (\textbf{Right panel}) Eventually, the source from the PBH evaporation is balanced by the diffusion and the system attains dynamical equilibrium.
        $r_\text{dec}$ represents the diffusion length at a given time [Eq.~\eqref{eq:rdec}]. $r_\text{cr}$ defines the critical radius within which the temperature is constant [see \textit{e.g.}, Eq.~\eqref{eq:rcr_ini}].
    }
	\label{fig:schematic}
\end{figure}

Once the high-energy particles are thermalized at $r \sim t_{\rm th}$, the thermal plasma is heated locally. The hot region then diffuses into the surrounding area (see the middle panel in Fig.~\ref{fig:schematic}).
The diffusion is a random process that is governed by the elastic scattering in the thermal plasma.
For a given time $t$, the diffusion length is estimated as
\begin{equation}
    r_\text{d} (t ,T(r)) \sim \sqrt{t_\text{el} t} \sim \sqrt{\frac{t}{\alpha^2 T(r)}},
\end{equation}
where $ T(r) $ is the local temperature of the plasma at radius $ r $,
and 
$t_\text{el}^{-1} \sim \alpha^2 T$ represents the interaction rate of the large-angle elastic scattering between particles in the thermal plasma.
The thermal diffusion defines two length scales which will turn out to be important in the following discussion.
The first one is the time scale of diffusion within a volume of $ 4\pi t_\text{th}^3 /3 $,
which is defined by $r_\text{d} (t_\text{d}, T) = t_\text{th}$, or equivalently
\begin{equation}
    t_\text{d} (M, T) \sim \alpha^{-2} T^{-2} \frac{\Mpl^2}{M},
\end{equation}
where we use $p \sim T_\text{H}$ and $T_\text{H} = \Mpl^2/M$.
For $t \gtrsim t_\text{d}$, the hot population initially at $r \sim t_\text{th}$ fills the inner region of $r \lesssim t_\text{th}$ and also propagates towards $r \gtrsim t_\text{th}$ by the thermal diffusion.
The other scale is the maximal length which the thermal diffusion can smooth within $t_\text{ev}$, denoted as $ r_{\rm dec} (M) $ (see the right panel of Fig.~\ref{fig:schematic}), which is defined as the solution of $ r_{\rm d} $ in the following equation 
\begin{equation}
    \label{eq:rdec}
    r_\text{d} (t_\text{ev} (M), T (r_\text{d})) \sim \alpha^{-1}\, \left( \frac{160}{\pi \mathcal{G} g_{H\ast} (T_\text{H})} \right)^{\frac{1}{2}}
    \left( \frac{M^3}{\Mpl^4 T (r_\text{d})} \right)^\frac{1}{2}.
\end{equation}
The plasma can reach local equilibrium within $ r_{\rm dec} $ while beyond that equilibrium cannot be obtained by diffusion.
Whenever a PBH with mass $M$ radiates its $\mathcal{O} (1)$ fraction of energy at $t \sim t_\text{ev} (M)$, $r_\text{dec} (M)$ decreases and the region $r \gtrsim r_\text{dec} (M)$ becomes decoupled from the system around a PBH.

\subsection{Formation and evolution of hot spots}
\label{sec:therm_proc}
Now we shall consider the detailed process of hot spot formation and evolution around a small PBH, taking into account the diffusion as well as thermalization of high-energy particles from Hawking radiation. 
In the following, we divide the evolution of hot spot into four regimes.

\subsubsection*{(i) \boldmath$t \lesssim t_\text{ev} (M_\text{ini})$}
During this period (i), the PBH mass can be approximated with $M \simeq M_\text{ini}$.
At the very early stage with $t \ll t_\text{ev} (M_\text{ini})$, the temperature of the ambient plasma is so high that $T > T_\text{H}$, where the Hawking radiation would not occur.
Owing to the cosmic expansion, the temperature of the ambient plasma eventually becomes so low that  $T_\text{H} > T$ because we have already shown $T_\text{H} > T_\text{ev}$ at least.
In the following discussion, we consider the later epoch where $T_\text{H} > T$ is already fulfilled.

For a moment, let us assume
$t_\text{d} (M_\text{ini}, T) \lesssim t_\text{ev} (M_\text{ini})$, or equivalently
\begin{equation}
    \label{eq:therm_i}
    \alpha^{-2}
    \left( \frac{\pi \mathcal{G} g_{H\ast} (T_\text{H})}{160} \right) \Mpl^6
    < 
    M_\text{ini}^4 T^2, 
\end{equation}
which is justified a posteriori.
In this case, we can approximate $M \simeq M_{\rm ini}$ within the time scales of thermalization and diffusion. 
At every time step in a unit of $t_\text{d} (M_\text{ini}, T)$, the high-energy particles with momenta $T_\text{H}$ get thermalized and forms a hot thermal plasma within $r \lesssim t_\text{th} (T_\text{H,ini},T)$, which implies
\begin{equation}
    \label{eq:homT_ini}
    \frac{\pi^2 g_\ast}{30} T^4 \times \frac{4 \pi}{3} t_\text{th}^3 (T_\text{H,ini},T) \simeq
    - t_\text{d} (M_\text{ini}, T) \left. \frac{\dd M}{\dd t} \right|_\text{ini} \quad
    \longrightarrow\quad
    T \simeq 2 \times 10^{-4}\, \left( \frac{\alpha}{0.1} \right)^\frac{8}{3} \left( \frac{g_\ast}{106.75} \right)^{-\frac{2}{3}}
    \left( \frac{g_{H\ast}}{108} \right)^\frac{2}{3} T_\text{H,ini}.
\end{equation}
Inserting this result back into Eq.~\eqref{eq:therm_i}, 
the above calculation is justified when 
\begin{equation}
    \label{eq:M*}
    M_\ast
    < 
    M_\text{ini},
    \qquad
    M_\ast \simeq 
    0.8 \,\mathrm{g}~ 
    \left( \frac{\alpha}{0.1} \right)^{-\frac{11}{3}}
    \left( \frac{g_\ast}{106.75} \right)^{\frac{2}{3}}
    \left( \frac{g_{H\ast}}{108} \right)^{-\frac{1}{6}}.
\end{equation}
One can see that the condition is almost always fulfilled in the parameter region of our interest; except for the tiny corner of the smallest possible PBH mass.

Finally, we estimate the temperature profile $T(r)$ around a PBH in Regime (i).
For a given time $t$, the region inside $r < r_\text{d} (t, T (r_\text{d}))$ is expected to be equilibrated by the efficient diffusion over that length scale.
In equilibrium, the total energy flux at any radius should be the same and is equal to the flux of Hawking radiation.
For a shell at $r < t_\text{th} (T_\text{H,ini},T)$, all the energy flux from the PBH is carried by the Hawking radiation itself, and hence the temperature inside $r < t_\text{th} (T_\text{H,ini},T)$ should not have the gradient but be homogeneous.
Note that this critical radius is given by 
\begin{equation}
    \label{eq:rcr_ini}
    r_\text{cr} (M_\text{ini}) \equiv
    t_\text{th} (T_\text{H,ini},T) \simeq 6 \times 10^7\, \left( \frac{\alpha}{0.1} \right)^{-6} \left( \frac{g_\ast}{106.75} \right)
    \left( \frac{g_{H\ast}}{108} \right)^{-1} T_\text{H,ini}^{-1},
\end{equation}
where we use Eq.~\eqref{eq:homT_ini} 
to estimate the homogeneous temperature of the core region $ r\lesssim t_{\rm th} $.
This is always much larger than the Schwarzschild radius $r_s \equiv 1/(4 \pi T_\text{H})$\footnote{
    This justfies that we use the Hawking temperature defined at the asymptotic infinity, which is not always the case near the Schwarzschild radius~\cite{Kim:2016iyf}.
}.
On the other hand, for the outer region of $r_\text{cr} (M_\text{ini}) < r$, high-energy particles from Hawking radiation already deposit their whole energy.
Thus, the temperature gradient should be responsible for the source of the energy flux.
Since the energy flux that leaves a shell at $r$ is estimated as
\begin{equation}
    \label{eq:energyflux}
    J_r \dd r \sim  t_\text{el} \left. T^4 \right|_{r} - t_\text{el} \left. T^4 \right|_{r + \dd r} 
    \quad
    \longrightarrow
    \quad
    \text{const.} = 4 \pi r^2 J_r \propto - r^2 \frac{\dd T^3}{\dd r},
\end{equation}
the temperature of the thermal plasma for $r > r_\text{cr} (M_\text{ini})$ should decrease in proportion to $r^{-1/3}$.

\begin{figure}[t]
	\centering
 	\includegraphics[width=0.33\linewidth]{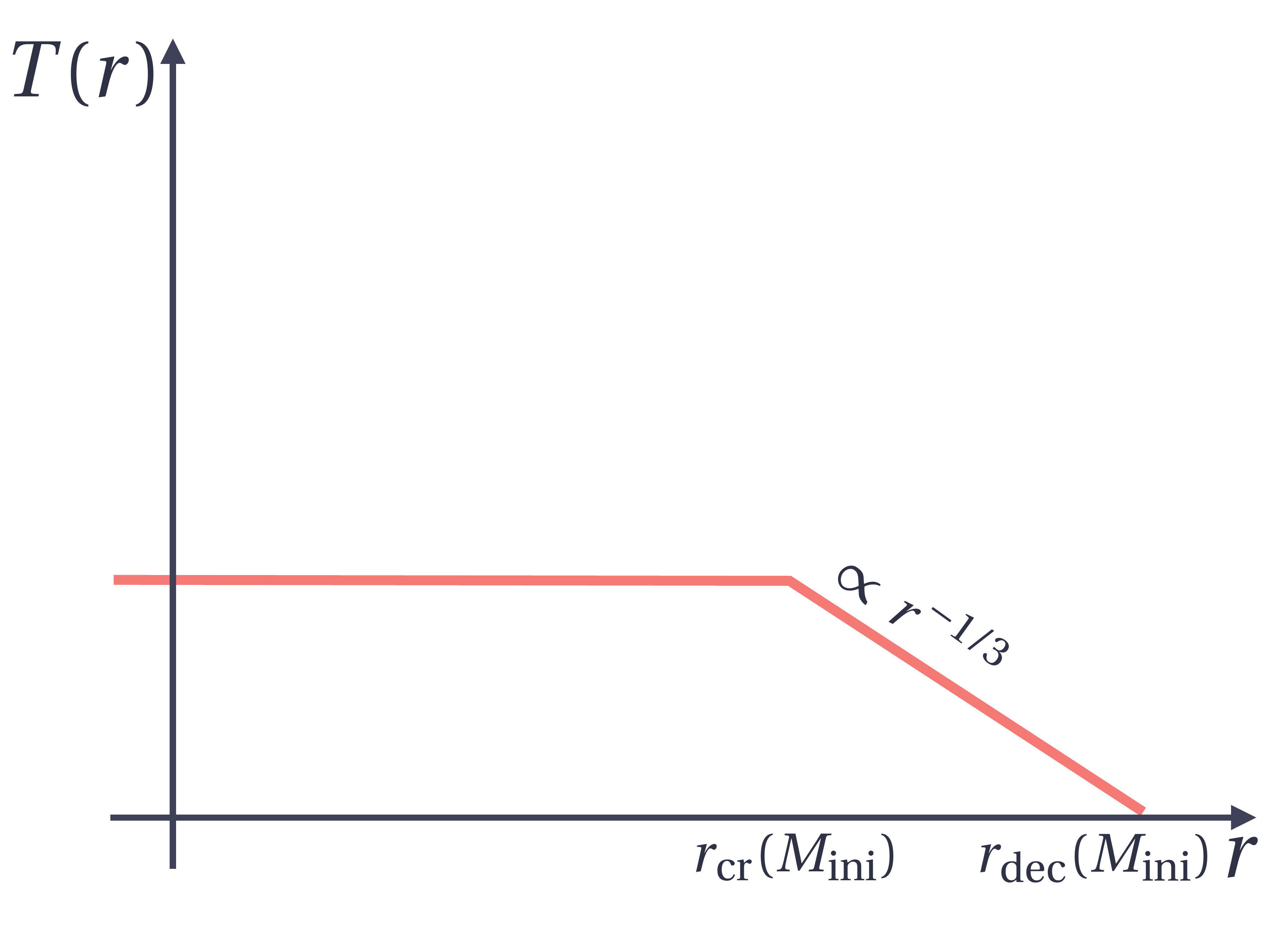}
    \includegraphics[width=0.33\linewidth]{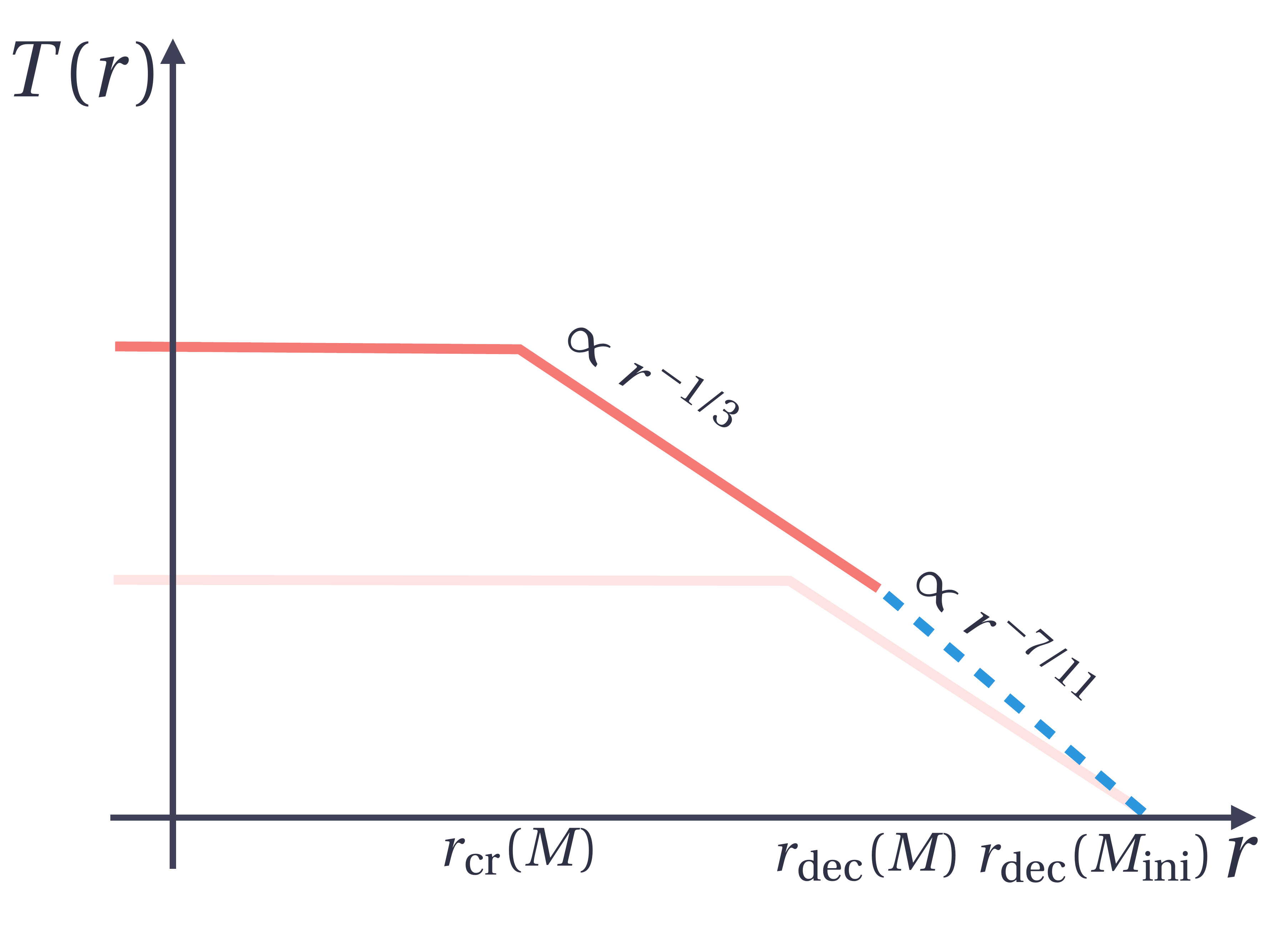} 
    \includegraphics[width=0.33\linewidth]{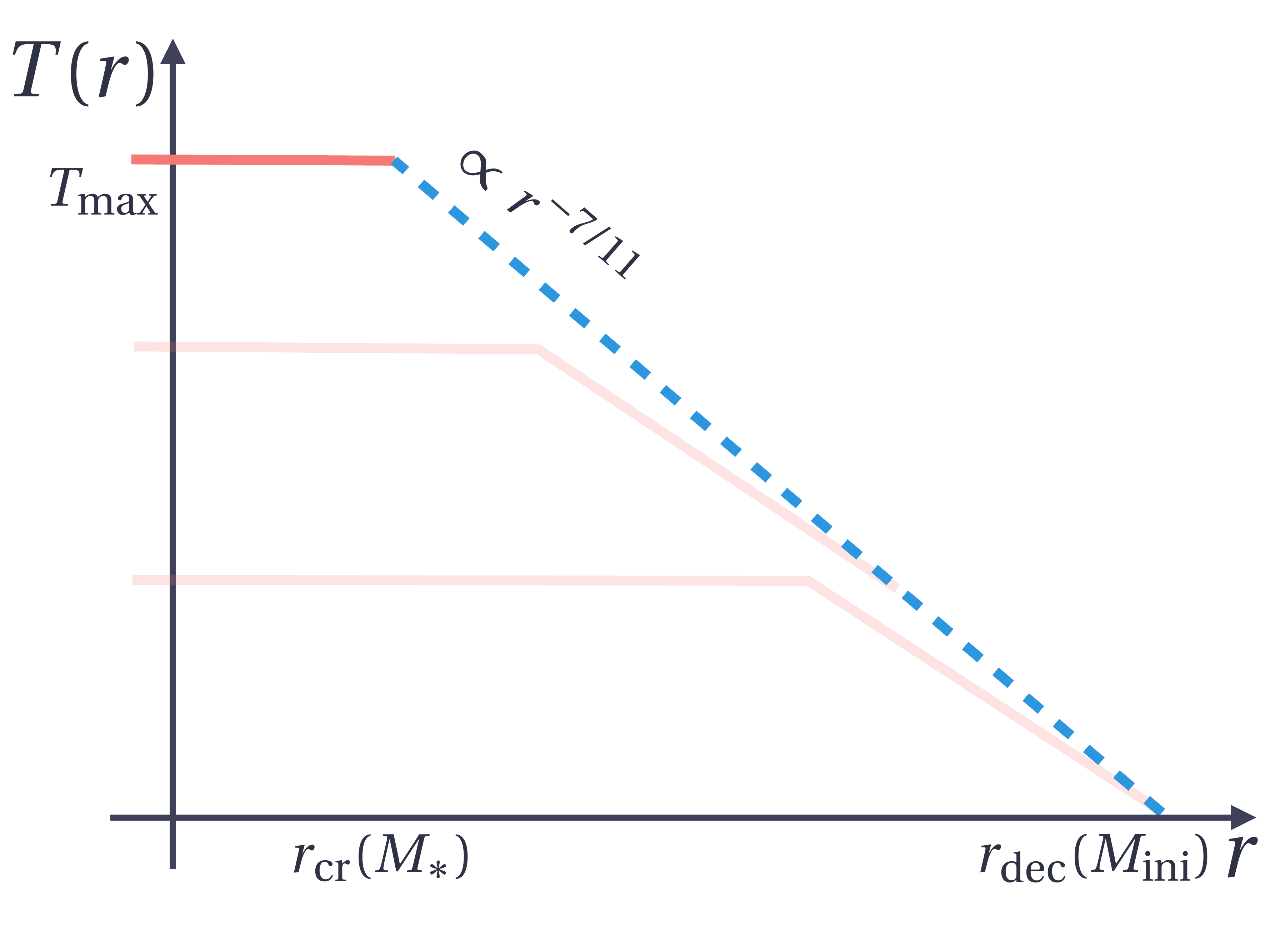} 
	\caption{
        The temperature profiles, $T(r)$, for Regimes (i), (ii), and (iii) are shown.
        The red solid lines represent the thermal plasma coupled to the inner population, and the blue dashed lines represent the decoupled thermal plasma.
        (\textbf{Left panel}) $T(r)$ at $t = t_\text{ev} (M_\text{ini})$ [Eq.~\eqref{eq:Tr_i}]. The critical radius, $r_\text{cr} (M_\text{ini})$, and the (maximal) diffusion length, $r_\text{dec} (M_\text{ini})$, are given in Eqs.~\eqref{eq:rcr_ini} and \eqref{eq:rdec_ini}, respectively.
        (\textbf{Middle panel}) $T(r)$ in Regime (ii) [Eqs.~\eqref{eq:Tr_ii_in} and \eqref{eq:Tr_ii_out}]. 
        The critical radius, $r_\text{cr} (M)$, and the (maximal) diffusion length, $r_\text{dec} (M)$, are given in Eqs.~\eqref{eq:rcr_ii} and \eqref{eq:rdec_ii}, respectively.
        Note that the PBH mass resides within $M_\ast < M < M_\text{ini}$.
        (\textbf{Right panel}) $T(r)$ in Regime (iii).
        As we discuss in the main text, the temperature profile does not change after the PBH mass becomes as small as $M_\ast$.
        The maximal temperature, $T_\text{max}$, is given in Eq.~\eqref{eq:Tmax}.
    }
	\label{fig:Tprofile}
\end{figure}

To sum up, the temperature profile reads
\begin{equation}
    \label{eq:Tr_i}
    T (r) \simeq 
    2 \times 10^{-4}\, \left( \frac{\alpha}{0.1} \right)^\frac{8}{3} \left( \frac{g_\ast}{106.75} \right)^{-\frac{2}{3}}
    \left( \frac{g_{H\ast}}{108} \right)^\frac{2}{3} T_\text{H,ini} \times
    \begin{cases}
        1 &\text{for} \quad r \lesssim r_\text{cr} (M_\text{ini}) ,\\
        \left( \frac{r_\text{cr} (M_\text{ini})}{r} \right)^{\frac{1}{3}}
        &\text{for} \quad r_\text{cr} (M_\text{ini}) \lesssim r,
    \end{cases}
\end{equation}
where the first line is from Eq.~\eqref{eq:homT_ini} while the second line simply connects the core region with the $ r^{-1/3} $ diffusion tail derived in the previous paragraph (see the left panel of Fig.~\ref{fig:Tprofile}).
Here, the scaling of $r^{-1/3}$ continues until the radius reaches the diffusion length scale within a given time $t$ or the local temperature around a PBH becomes smaller than the background temperature.
The maximal diffusion length is 
\begin{equation}
    \label{eq:rdec_ini}
    r_\text{dec} (M_\text{ini}) 
    \simeq 3 \times 10^{-10} \, \mathrm{sec}~ \left( \frac{\alpha}{0.1} \right)^{-\frac{8}{5}}
    \left( \frac{g_\ast}{106.75} \right)^{\frac{1}{5}} 
    \left( \frac{g_{H_\ast}}{108} \right)^{-\frac{4}{5}}
    \left( \frac{T_\text{H,ini}}{10^4 \,\mathrm{GeV}} \right)^{-\frac{11}{5}}.
\end{equation}
One obtains the (would-be) temperature at $r_\text{dec} (M_\text{ini})$ by inserting it into Eq.~\eqref{eq:Tr_i}:
\begin{equation}
    \label{eq:Trdec_ini}
    T (r_\text{dec} (M_\text{ini})) \sim 0.4 \,\mathrm{MeV} ~
    \left( \frac{\alpha}{0.1} \right)^{\frac{6}{5}} 
    \left( \frac{g_\ast}{106.75} \right)^{-\frac{2}{5}}
    \left( \frac{g_{H\ast}}{108} \right)^{\frac{3}{5}}
    \left( \frac{T_\text{H,ini}}{10^4 \,\mathrm{GeV}} \right)^{\frac{7}{5}}.
\end{equation}
As we have $T_\text{ev} \propto M_\text{ini}^{-3/2}$, the background temperature exceeds $T (r_\text{dec})$ for a sufficiently small $M_\text{ini}$.

\subsubsection*{(ii) \boldmath$M_\ast \lesssim M \lesssim M_\text{ini}$}
After $t \gtrsim t_\text{ev} (M_\text{ini})$, the PBH mass changes significantly by emitting the Hawking radiation.
For a given mass $M$, the typical time scale to lose an $\mathcal{O}(1)$ amount of its mass is given by $t_\text{ev} (M)$.
Although the PBH mass becomes significantly time-dependent contrary to Regime (i), the thermalization is efficient in a shorter time scale as far as the condition, $t_\text{d} (M, T) < t_\text{ev} (M)$, is met, which reads $M_\ast < M$.
Therefore, the analysis in Regime (i) can be applied to Regime (ii) immediately once we replace $M_\text{ini}$ with $M$.

Let us just summarize the results in the following because the derivation is essentially the same as the previous case (i).
The temperature profile up to $r < r_\text{dec} (M)$ is given by
\begin{equation}
    \label{eq:Tr_ii_in}
    T (r) \sim 2 \times 10^{-4}\, \left( \frac{\alpha}{0.1} \right)^\frac{8}{3} \left( \frac{g_\ast}{106.75} \right)^{-\frac{2}{3}}
    \left( \frac{g_{H\ast}}{108} \right)^\frac{2}{3} T_\text{H} (M) \times
    \begin{cases}
        1 &\text{for} \quad r \lesssim r_\text{cr} (M)\\
        \left( \frac{r_\text{cr} (M)}{r} \right)^{\frac{1}{3}}
        &\text{for} \quad r_\text{cr} (M) \lesssim r \lesssim r_\text{dec} (M),
    \end{cases}
\end{equation}
where
\begin{align}
    \label{eq:rcr_ii}
    r_\text{cr} (M) &\simeq 6 \times 10^7\, \left( \frac{\alpha}{0.1} \right)^{-6} \left( \frac{g_\ast}{106.75} \right)
    \left( \frac{g_{H\ast}}{108} \right)^{-1} T_\text{H}^{-1} (M), \\
    \label{eq:rdec_ii}
    r_\text{dec} (M) & \simeq 3 \times 10^{-10} \, \mathrm{sec}~ \left( \frac{\alpha}{0.1} \right)^{-\frac{8}{5}}
    \left( \frac{g_\ast}{106.75} \right)^{\frac{1}{5}} 
    \left( \frac{g_{H_\ast}}{108} \right)^{-\frac{4}{5}}
    \left( \frac{T_\text{H} (M)}{10^4 \,\mathrm{GeV}} \right)^{-\frac{11}{5}}.
\end{align}
As the PBH mass decreases, the overall height of $T(r)$ increases in proportion to $M^{-1}$ while the critical radius and the diffusion length decrease as $M$ and $M^{11/5}$, respectively.
This implies that the region of $r_\text{dec} (M) \lesssim r \lesssim r_\text{dec} (M_\text{ini})$ initially resides within the diffusion length but is decoupled at a later epoch. 
Once it is decoupled, such a hot region remains as the following envelope,
\begin{equation}
    \label{eq:Tr_ii_out}
    T(r) \simeq 
    0.4 \,\mathrm{MeV} ~
    \left( \frac{\alpha}{0.1} \right)^{\frac{6}{5}} 
    \left( \frac{g_\ast}{106.75} \right)^{-\frac{2}{5}}
    \left( \frac{g_{H\ast}}{108} \right)^{\frac{3}{5}}
    \left( \frac{T_\text{H,ini}}{10^4 \,\mathrm{GeV}} \right)^{\frac{7}{5}}
    \left( \frac{r_\text{dec} (M_\text{ini})}{r} \right)^{\frac{7}{11}}
    \quad \text{for} \quad r_\text{dec} (M) \lesssim r \lesssim r_\text{dec} (M_\text{ini}),
\end{equation}
where we use 
Eqs.~\eqref{eq:rdec_ini} and \eqref{eq:Trdec_ini} with $ M_{\rm ini} $ replaced by $ M $.
See also Fig.~\ref{fig:Tprofile} for intuitive illustration.

When the PBH mass becomes as small as $M_\ast$ given in Eq.~\eqref{eq:M*}, the two scales merge by definition, \textit{i.e.}, $r_\text{cr} (M_\ast) = r_\text{dec} (M_\ast)$,
and the highest temperature inside $r \lesssim r_\text{cr} (M_\ast)$ is achieved, which is given by
\begin{align}
    \label{eq:Tmax}
    T_\text{max} \equiv \left. T (r < r_\text{cr}) \right|_{M_\ast} &\simeq 0.02 ~ \alpha^{19/3} g_\ast^{-4/3} g_{H\ast}^{5/6}~ M_{\rm pl} ~, \\
    &\simeq 2 \times 10^9 \, \mathrm{GeV}~
    \left( \frac{\alpha}{0.1} \right)^{\frac{19}{3}} 
    \left( \frac{g_\ast}{106.75} \right)^{- \frac{4}{3}}
    \left( \frac{g_{H\ast}}{108} \right)^{\frac{5}{6}}.
\end{align}
As we will see shortly, $T_\text{max}$ is indeed the maximal temperature realized around evaporating PBHs.
Remarkably enough, $T_\text{max}$ never depends on the initial PBH masses.

\subsubsection*{(iii) \boldmath$ 0 < M \lesssim M_\ast$}
Once the PBH mass becomes smaller than $M_\ast$, the diffusion cannot catch up with the evaporation time scale, \textit{i.e.}, $t_\text{d} (M) \gtrsim t_\text{ev} (M)$.
Even if the high-energy particles deposit their energy at $r \gtrsim r_\text{cr} (M_\ast)$, 
the heated shell does not have enough time to diffuse and just remains as it is. 
This is one of the most important regimes in which the finiteness of the thermalization time scale $t_{\rm th}$ significantly affects the formation of hot spots around PBHs. 

As the PBH mass decreases, the Hawking temperature increases in proportion to $M^{-1}$.
However, 
contrary to the earlier Regimes (i) and (ii), 
heating thermal plasma inside $r < t_\text{th}$ is not accompanied due to the inefficiency of the diffusion mentioned above. 
The high-energy particles emitted in this regime can deposit its energy only at $r \gtrsim r_\text{cr} (M_\ast)$ because its thermalization time scale grows in proportion to $M^{-1/2}$ at least.
Since the total energy emitted after $t > t_\text{ev} (M_*) $ is always smaller than that emitted prior to $t_\text{ev}(M_*) $, the corresponding energy density cannot exceed that produced in the earlier regime unless the high-energy particles lose their energy within a smaller radius, which, however, is not the case as we have discussed.
Therefore, the energy deposited in Regime (iii) cannot heat up the hot spot further. 
As a result, 
the resulting maximal temperature of the hot spot remains as in Eq.~\eqref{eq:Tmax}. 
The temperature profile in this regime is therefore given by 
\begin{equation}
    \label{eq:mainresult}
    T (r) \sim T_{\rm max}
    \begin{cases}
        1 &\text{for} \quad r \lesssim r_\text{cr} (M_*),\\
        \left( \frac{r_\text{cr} (M_*)}{r} \right)^{\frac{7}{11}}
        &\text{for} \quad r_\text{cr} (M_*) \lesssim r \lesssim r_\text{dec} (M_{\rm ini}),
    \end{cases}
\end{equation}
with 
\begin{equation}
    r_\text{cr} (M_*) \simeq 3 \times 10^{-30} \, {\rm sec}\, \left( \frac{\alpha}{0.1} \right)^{-\frac{29}{3}} \left( \frac{g_\ast}{106.75} \right)^{ \frac{5}{3}}
    \left( \frac{g_{H\ast}}{108} \right)^{-\frac{7}{6}} .
\end{equation}
This result
is significantly different from the one estimated in Ref.~\cite{Das:2021wei} under the instantaneous thermalization approximation. Note that $ r_{\rm cr}(M_*) \gg r_s $ always holds.

\subsubsection*{(iv) Cooling after evaporation}

After the completion of the PBH evaporation, the high-temperature region starts to cool down as the energy dissipates. The analysis of the cooling process is shown in Ref.~\cite{Das:2021wei} which we briefly summarize here. As cooling is also the dissipation of energy, it obeys the same physics as the heating process. The diffusion equation is given as 
\begin{equation}
    \frac{\partial \rho}{\partial t} = - \frac{\partial}{\partial r} J_r \propto \frac{\partial}{\partial r} \left( \frac{\partial T^3}{\partial r} \right) ,
\end{equation}
where $ \rho \propto T^4 $ and we have used Eq.~\eqref{eq:energyflux} in the proportionality. For $ r > r_{\rm cr} (M_*) $, we have $ T\propto r^{-7/11} $ as shown previously and it is easy to see that the typical time scale of cooling is 
\begin{equation}
    t_{\rm cool} \propto r^{15/11} ,
\end{equation}
which means that inner part of the hot spot cools faster. As a result, the outer region remains almost unchanged while the inner part gradually cools down. Since $ T (r) = {\rm const.} $ for $ r< r_{\rm cr} (M_*) $, the core transfers its energy to the outter region as a whole, so the temperature of the core decreases while its radius increases.

\section{Implications for phenomenology}
\label{sec:implication}


In this section, we briefly discuss some phenomenological implications of hot spots around evaporating PBHs to cosmological scenarios.

\subsection{Sphaleron conversion and baryon asymmetry}

Since the temperature around small PBHs can exceed the electroweak scale as shown previously, 
the electroweak symmetry can be restored and the sphaleron process can become efficient around the PBHs. 
This leads to an interesting possibility for baryogenesis 
if the lepton asymmetry is generated after the electroweak phase transition. 
For example, 
if one introduces sterile neutrinos with masses of 
${\cal O}(10^{-2})\,\mathrm{eV}$ -- ${\cal O}(1)\,\mathrm{eV}$ scales, 
lepton asymmetry can be produced via the
oscillation between active and sterile neutrinos at around cosmic temperature of the order ${\cal O}(1)\, \mathrm{MeV}$ -- ${\cal O}(10)\,\mathrm{MeV}$~\cite{Shi:1996ic,Shi:1999kg} (see also a review article~\cite{Dolgov:2002wy}). 
The Affleck-Dine mechanism also provides a scenario to produce  large lepton (flavor) asymmetry which can be protected by Q-balls down to a much lower temperature than the electroweak scale~\cite{Kawasaki:2002hq}. 
The existence of such a large lepton flavor asymmetry is motivated to reduce the $^4$He abundance~\cite{Kohri:1996ke} and solve the anomaly reported by the EMPRESS VIII collaboration~\cite{Matsumoto:2022tlr}.

Suppose that lepton asymmetry is generated at a low temperature and estimate how much baryon asymmetry is generated around the PBHs. 
%
We first need to calculate the volume fraction in which the electroweak symmetry is restored after the evaporation of PBHs. Let us assume the volume fraction of the region with temperature higher than some temperature $ T_1 $ at the time of evaporation such that $ T_{\rm max} > T_1 > T_{\rm ev} $ is given by $ f_1 $. Given the temperature profile of the ambient plasma around PBHs at evaporation \eqref{eq:mainresult}, one can obtain 
\begin{equation}
	r_1 = \left( \frac{T_{\rm max}}{T_1} \right)^{\frac{11}{7}} r_{\rm cr} (M_*) ~,
\end{equation}
where $  r_1 $ denotes the distance between the shell with temperature $ T_1 $ and the PBH. As the mean separation between PBHs at evaporation is shown in Eq.~\eqref{eq:distance}, the volume fraction, $ f_1 $, of the region with $ T>T_1 $ can be estimated as 
\begin{align}
	f_1 (T_1) &\sim \frac{4 \pi}{3} \left(\frac{r_1}{L_{\rm PBH, ev}}\right)^3 \nonumber \\
	&\sim  
    6 \times 10^{-60} \, \left( \frac{2\times 10^9 ~{\rm GeV}}{T_1} \right)^{\frac{33}{7}} {\rm Min}\left[ 1, \beta \frac{T_{\rm ini}}{T_{\rm ev}} \right] \left( \frac{\alpha}{0.1} \right)^{\frac{6}{7}} \left( \frac{g_*}{106.75} \right)^{-\frac{9}{7}} \left( \frac{g_{H*}}{108} \right)^{\frac{17}{7}} \left( \frac{M_{\rm ini}}{10^9 \, \mathrm{g}} \right)^{-7} ~,
\end{align}
which highly depends on the initial mass of PBHs as the mean separation depends on $ M_{\rm ini} $. As the first example, we estimate the volume fraction with temperature higher than reheating temperature, assuming that PBHs once dominate the Universe and complete reheating by Hawking radiation at the moment of evaporation. In this case, the reheating temperature is given in Eq.~\eqref{eq:Tev}. Therefore, one can obtain the volume fraction with temperature higher than $ T_{\rm rh} $ as 
\begin{equation}
	f_1 (T_{\rm rh}) \sim 
    0.3 \, {\rm Min}\left[ 1, \beta \frac{T_{\rm ini}}{T_{\rm ev}} \right] \left( \frac{\alpha}{0.1} \right)^{\frac{6}{7}} \left( \frac{g_*}{106.75} \right)^{-\frac{3}{28}} \left( \frac{g_{H*}}{108} \right)^{\frac{1}{14}} \left( \frac{M_{\rm ini}}{10^9 \, \mathrm{g}} \right)^{\frac{1}{14}} ~.
\end{equation}
One can also estimate the volume fraction for the region with $ T $ higher than the decoupling temperature of electroweak sphaleron $ \sim 132 ~{\rm GeV} $~\cite{DOnofrio:2014rug}, which is essential to see whether these regions have the possibility to realize baryogenesis. For $ T_1 = T_{\rm sph} = 132 ~{\rm GeV} $, it is easy to find that 
\begin{equation}
    f_{\rm sph} (T_{\rm sph}) \sim 
    0.1 ~ {\rm Min}\left[ 1, \beta \frac{T_{\rm ini}}{T_{\rm ev}} \right] \left( \frac{\alpha}{0.1} \right)^{\frac{6}{7}} \left( \frac{g_*}{106.75} \right)^{-\frac{9}{7}} \left( \frac{g_{H*}}{108} \right)^{\frac{17}{7}} \left( \frac{M_{\rm ini}}{10^{5.5} \, \mathrm{g}} \right)^{-7} ~,
\end{equation}
which can be close to $ \mathcal{O}(1) $ because of the strong dependence of the volume fraction on the initial PBH mass. However, $ M_{\rm ini} $ cannot be smaller than $ 10^{5.5} {\rm g} $ because this estimation is meaningful only when $ T_{\rm ev} < T_{\rm sph} $. 

\subsection{Symmetry restoration and formation of topological defects}

One may consider the physics beyond the Standard Model with spontaneously broken symmetries. 
In general, symmetry is restored at a high temperature because the thermal effect on Higgs potential tends to favor the symmetry-enhanced point in the phase space. 
As the temperature decreases, the symmetry is spontaneously broken by the Higgs mechanism. 
Depending on the topology of the vacuum structure, topological defects, such as cosmic strings, domain walls, and monopoles, may form after the spontaneous symmetry breaking. 
Let us consider a model with those properties without specify the detail of the model.

Suppose that the maximal temperature \eqref{eq:Tmax} exceeds the critical temperature of some symmetry. 
Then the symmetry is restored within the hot spots around the evaporating PBHs. 
Subsequently, as the temperature cools down, 
the symmetry is again spontaneously broken 
and the topological defects may form if the correlation length at the phase transition is shorter than the size of the symmetry-restored region. 
However, 
the hot spots are located only around the PBHs 
and therefore the topological defects form locally. 
This particularly means that longer strings or larger domain walls than the mean separation length of PBHs cannot be produced. 
As a result, cosmic strings and domain walls disappear soon after the formation because they tend to be contracted to a point by their tensions. 
On the contrary, monopoles can survive if their annihilation can be neglected. 
Fortunately, 
the maximal temperature around the PBH does not exceed the grand unified theory (GUT) scale of order $10^{16}$ GeV, 
so that we do not expect the production of monopoles in a typical unified theory of SM gauge couplings. 
Still, one may consider a model with a hidden sector which predicts a hidden monopole at a lower scale than \eqref{eq:Tmax}. 
In this case, the monopole production in the hot spots leads to an interesting cosmological possibility because, \textit{e.g.}, it is a good candidate for dark matter.

One may also consider a model in which the present vacuum is a false vacuum 
and there is a true vacuum at a large field value of some scalar field. 
It is possible that the true vacuum bubbles are created in the hot spot around the evaporating PBHs. 
In this case, the true vacuum bubbles tend to expand after the formation and the Universe may be dominated by the true vacuum state. 
This could be the case for the SM Higgs, where the electroweak vacuum is a false vacuum and there is a true vacuum at a large field value. 
However, one has to take into account the thermal corrections to the Higgs potential, which stabilize the Higgs field at the origin of the potential even at a very high-temperature. 
Thanks to this effect, 
we expect that the Higgs vacuum decay may not happen in the hot spots.

The catalyzed vacuum decay around the PBHs was discussed in Refs.~\cite{Gregory:2013hja,Burda:2015isa,Burda:2015yfa,Burda:2016mou,Tetradis:2016vqb,Canko:2017ebb} by using the Euclidean metric. 
In Refs.~\cite{Mukaida:2017bgd,Kohri:2017ybt}, we discuss that the effects discussed in those papers 
would be due to the thermal effects, 
which fills in the whole Universe as a consequence of the Euclidean metric (see also Refs.~\cite{Gorbunov:2017fhq,Shkerin:2021rhy,Ai:2022kqm,Strumia:2022jil}). 
The symmetry restoration in the hot spots discussed in the present paper is similar but different from those effects. 
Since the hot spots are generated by the interactions between the Hawking radiation and the ambient plasma, they have nothing to do with Euclideanization and can be formed in a more realistic situation where the Hawking temperature is much larger than that of the ambient plasma.
In this sense, the vacuum decay in the hot spots is nothing but the ordinary thermal transition. 

\subsection{Dark matter production}

Finally, we briefly comment on the weakly interacting massive particle (WIMP) production in the hot spots. 
If one considers a model with a WIMP, it can be produced at a high temperature plasma around the PBHs. 
However, we expect that the DM can also be directly produced from the evaporating PBHs as the Hawking radiation. 
Moreover, the DM can also be produced non-thermally during the thermalization process, which was discussed in detail in the context of pre-thermal phase of reheating of the Universe~\cite{Harigaya:2014waa,Harigaya:2019tzu,Drees:2021lbm,Drees:2022vvn,Mukaida:2022bbo}. 
One may have to specify the DM model and calculate all those contributions to estimate the resulting DM abundance. 
We leave the detailed discussion about this for a future work.

\section{Summary}
\label{sec:conclusion}

The temperature of the Hawking radiation characterizes its ``thermal'' spectrum but does not refer to the thermal equilibrium with the surrounding plasma. Because the typical energy of the Hawking radiation can be as large as the Planck scale, it is non-trivial how those high-energy particles are thermalized and heat up the ambient plasma. 
In this paper, we have investigated the thermalization process of the Hawking radiation emitted by PBHs in the early Universe and shown the resulting temperature profile of the plasma around a PBH. 
We have extended the analysis in Ref.~\cite{Das:2021wei} by taking into account a finite time scale of thermalization of high-energy particles in low-temperature plasma. 
In particular, the thermalization rate of higher-energy particles is more suppressed by the LPM effect and hence the Hawking radiation from a small PBH may not be able to heat up the region in the vicinity of PBH. 
This effect together with diffusion generates an inner core region surrounding the PBHs. 
As the PBH mass decreases, 
the temperature of the inner core increases and eventually reaches 
the maximal temperature 
of Eq.~\eqref{eq:Tmax}. 
The temperature profile in the outer region is determined mainly by diffusion as shown in the previous work. 
We summarize the process as follows (also see Fig.~\ref{fig:Tprofile}). 
\begin{itemize}
    \item At early time $ t\lesssim t_{\rm ev} (M_{\rm ini}) $, the Hawking radiation does not significantly changes the PBH mass. The high-energy particles from the PBHs get thermalized after $ t_{\rm th} $ which is determined by the LPM effect. The energy deposited by these particles dissipates over a region $ r\sim t_{\rm th} $ with a time scale much shorter than the PBH lifetime, creating an inner core and a $ r^{-1/3} $ tail in the outer region. 
    \item At a later time, the time-dependence of the PBH masses is no longer negligible. 
    In the regime of $ M_* \lesssim M \lesssim M_{\rm ini} $, where $M_*$ is given by \eqref{eq:M*}, 
    the core temperature increases whereas the core size decreases. Meanwhile, the dissipation 
    length 
    also decreases, 
    leaving a decoupled tail with a power of $ r^{-7/11} $ for a large $r$. 
    The core finally reaches its maximal temperature of \eqref{eq:Tmax} when the diffusion cannot catch up with the evaporation time scale 
    at $ M= M_* $. 
    The maximal temperature is particularly independent of the initial PBH mass. 
    \item When the PBH mass becomes smaller than $M_* $, the Hawking radiation has too large energy to be thermalized and dissipated in the vicinity of the PBH. 
    Because the energy of Hawking radiation is not deposited into the plasma near the PBH, the core temperature does not change in this regime. Since the total energy left in the PBH is small, it cannot affect the temperature profile of the outer region either. 
\end{itemize}

The formation of hot spots around PBHs in the early Universe may have interesting implications for phenomenology such as baryogenesis, catalysis of vacuum decay, and DM production. 
In particular, 
in contrast to the previous estimation in Ref.~\cite{Das:2021wei}, 
we have found that 
the formation of extended objects such as GUT monopoles in the heated plasma may not be possible 
because the maximal temperature is much lower than the GUT scale.

\section*{Acknowledgments}
M.\,H.\, would like to thank Bardia Najjari for the useful discussion at the 27th PASCOS in Heidelberg. 
K.\,K.\, was supported by JSPS KAKENHI Grant No.\ JP17H01131, and by MEXT KAKENHI Grant Nos.\ JP20H04750 and JP22H05270.
K.\,M.\, was supported by MEXT Leading Initiative for Excellent Young Researchers Grant No.\ JPMXS0320200430,
and by JSPS KAKENHI Grant No.\ 	JP22K14044.
M.\,Y.\, was supported by MEXT Leading Initiative for Excellent Young Researchers, and by JSPS KAKENHI Grant No.\ JP20H05851 and JP21K13910.

\small
\bibliographystyle{utphys}
\bibliography{ref}

\providecommand{\href}[2]{#2}\begingroup\raggedright\begin{thebibliography}{10}

\bibitem{Hawking:1974rv}
S.~W. Hawking, ``{Black hole explosions},''
  \href{http://dx.doi.org/10.1038/248030a0}{{\em Nature} {\bfseries 248} (1974)
  30--31}.

\bibitem{Hawking:1975vcx}
S.~W. Hawking, ``{Particle Creation by Black Holes},''
  \href{http://dx.doi.org/10.1007/BF02345020}{{\em Commun. Math. Phys.}
  {\bfseries 43} (1975) 199--220}. [Erratum: Commun.Math.Phys. 46, 206 (1976)].

\bibitem{Zeldovich:1967lct}
Y.~B. Zel'dovich and I.~D. Novikov, ``{The Hypothesis of Cores Retarded during
  Expansion and the Hot Cosmological Model},'' {\em Soviet Astron. AJ (Engl.
  Transl. ),} {\bfseries 10} (1967) 602.

\bibitem{Hawking:1971ei}
S.~Hawking, ``{Gravitationally collapsed objects of very low mass},'' {\em Mon.
  Not. Roy. Astron. Soc.} {\bfseries 152} (1971) 75.

\bibitem{Carr:1974nx}
B.~J. Carr and S.~W. Hawking, ``{Black holes in the early Universe},'' {\em
  Mon. Not. Roy. Astron. Soc.} {\bfseries 168} (1974) 399--415.

\bibitem{Carr:1975qj}
B.~J. Carr, ``{The Primordial black hole mass spectrum},''
  \href{http://dx.doi.org/10.1086/153853}{{\em Astrophys. J.} {\bfseries 201}
  (1975) 1--19}.

\bibitem{Garriga:2015fdk}
J.~Garriga, A.~Vilenkin, and J.~Zhang, ``{Black holes and the multiverse},''
  \href{http://dx.doi.org/10.1088/1475-7516/2016/02/064}{{\em JCAP} {\bfseries
  02} (2016) 064}, \href{http://arxiv.org/abs/1512.01819}{{\ttfamily
  arXiv:1512.01819 [hep-th]}}.

\bibitem{Deng:2017uwc}
H.~Deng and A.~Vilenkin, ``{Primordial black hole formation by vacuum
  bubbles},'' \href{http://dx.doi.org/10.1088/1475-7516/2017/12/044}{{\em JCAP}
  {\bfseries 12} (2017) 044}, \href{http://arxiv.org/abs/1710.02865}{{\ttfamily
  arXiv:1710.02865 [gr-qc]}}.

\bibitem{Deng:2018cxb}
H.~Deng, A.~Vilenkin, and M.~Yamada, ``{CMB spectral distortions from black
  holes formed by vacuum bubbles},''
  \href{http://dx.doi.org/10.1088/1475-7516/2018/07/059}{{\em JCAP} {\bfseries
  07} (2018) 059}, \href{http://arxiv.org/abs/1804.10059}{{\ttfamily
  arXiv:1804.10059 [gr-qc]}}.

\bibitem{Carr:2020gox}
B.~Carr, K.~Kohri, Y.~Sendouda, and J.~Yokoyama, ``{Constraints on primordial
  black holes},'' \href{http://dx.doi.org/10.1088/1361-6633/ac1e31}{{\em Rept.
  Prog. Phys.} {\bfseries 84} no.~11, (2021) 116902},
  \href{http://arxiv.org/abs/2002.12778}{{\ttfamily arXiv:2002.12778
  [astro-ph.CO]}}.

\bibitem{Chapline:1975ojl}
G.~F. Chapline, ``{Cosmological effects of primordial black holes},''
  \href{http://dx.doi.org/10.1038/253251a0}{{\em Nature} {\bfseries 253}
  no.~5489, (1975) 251--252}.

\bibitem{LIGOScientific:2016aoc}
{\bfseries LIGO Scientific, Virgo} Collaboration, B.~P. Abbott {\em et~al.},
  ``{Observation of Gravitational Waves from a Binary Black Hole Merger},''
  \href{http://dx.doi.org/10.1103/PhysRevLett.116.061102}{{\em Phys. Rev.
  Lett.} {\bfseries 116} no.~6, (2016) 061102},
  \href{http://arxiv.org/abs/1602.03837}{{\ttfamily arXiv:1602.03837 [gr-qc]}}.

\bibitem{Sasaki:2016jop}
M.~Sasaki, T.~Suyama, T.~Tanaka, and S.~Yokoyama, ``{Primordial Black Hole
  Scenario for the Gravitational-Wave Event GW150914},''
  \href{http://dx.doi.org/10.1103/PhysRevLett.117.061101}{{\em Phys. Rev.
  Lett.} {\bfseries 117} no.~6, (2016) 061101},
  \href{http://arxiv.org/abs/1603.08338}{{\ttfamily arXiv:1603.08338
  [astro-ph.CO]}}. [Erratum: Phys.Rev.Lett. 121, 059901 (2018)].

\bibitem{Hooper:2020evu}
D.~Hooper, G.~Krnjaic, J.~March-Russell, S.~D. McDermott, and
  R.~Petrossian-Byrne, ``{Hot Gravitons and Gravitational Waves From Kerr Black
  Holes in the Early Universe},''
  \href{http://arxiv.org/abs/2004.00618}{{\ttfamily arXiv:2004.00618
  [astro-ph.CO]}}.

\bibitem{Carr:1976zz}
B.~J. Carr, ``{Some cosmological consequences of primordial black-hole
  evaporations},'' \href{http://dx.doi.org/10.1086/154351}{{\em Astrophys. J.}
  {\bfseries 206} (1976) 8--25}.

\bibitem{Barrow:1981zv}
J.~D. Barrow and G.~G. Ross, ``{Cosmological Constraints on the Scale of Grand
  Unification},'' \href{http://dx.doi.org/10.1016/0550-3213(81)90536-8}{{\em
  Nucl. Phys. B} {\bfseries 181} (1981) 461--486}.

\bibitem{Barrow:1990he}
J.~D. Barrow, E.~J. Copeland, E.~W. Kolb, and A.~R. Liddle, ``{Baryogenesis in
  extended inflation. 2. Baryogenesis via primordial black holes},''
  \href{http://dx.doi.org/10.1103/PhysRevD.43.984}{{\em Phys. Rev. D}
  {\bfseries 43} (1991) 984--994}.

\bibitem{Baumann:2007yr}
D.~Baumann, P.~J. Steinhardt, and N.~Turok, ``{Primordial Black Hole
  Baryogenesis},'' \href{http://arxiv.org/abs/hep-th/0703250}{{\ttfamily
  arXiv:hep-th/0703250}}.

\bibitem{Fujita:2014hha}
T.~Fujita, M.~Kawasaki, K.~Harigaya, and R.~Matsuda, ``{Baryon asymmetry, dark
  matter, and density perturbation from primordial black holes},''
  \href{http://dx.doi.org/10.1103/PhysRevD.89.103501}{{\em Phys. Rev. D}
  {\bfseries 89} no.~10, (2014) 103501},
  \href{http://arxiv.org/abs/1401.1909}{{\ttfamily arXiv:1401.1909
  [astro-ph.CO]}}.

\bibitem{Hook:2014mla}
A.~Hook, ``{Baryogenesis from Hawking Radiation},''
  \href{http://dx.doi.org/10.1103/PhysRevD.90.083535}{{\em Phys. Rev. D}
  {\bfseries 90} no.~8, (2014) 083535},
  \href{http://arxiv.org/abs/1404.0113}{{\ttfamily arXiv:1404.0113 [hep-ph]}}.

\bibitem{Hamada:2016jnq}
Y.~Hamada and S.~Iso, ``{Baryon asymmetry from primordial black holes},''
  \href{http://dx.doi.org/10.1093/ptep/ptx011}{{\em PTEP} {\bfseries 2017}
  no.~3, (2017) 033B02}, \href{http://arxiv.org/abs/1610.02586}{{\ttfamily
  arXiv:1610.02586 [hep-ph]}}.

\bibitem{Bernal:2021yyb}
N.~Bernal, F.~Hajkarim, and Y.~Xu, ``{Axion Dark Matter in the Time of
  Primordial Black Holes},''
  \href{http://dx.doi.org/10.1103/PhysRevD.104.075007}{{\em Phys. Rev. D}
  {\bfseries 104} (2021) 075007},
  \href{http://arxiv.org/abs/2107.13575}{{\ttfamily arXiv:2107.13575
  [hep-ph]}}.

\bibitem{Bernal:2021bbv}
N.~Bernal, Y.~F. Perez-Gonzalez, Y.~Xu, and O.~Zapata, ``{ALP dark matter in a
  primordial black hole dominated universe},''
  \href{http://dx.doi.org/10.1103/PhysRevD.104.123536}{{\em Phys. Rev. D}
  {\bfseries 104} no.~12, (2021) 123536},
  \href{http://arxiv.org/abs/2110.04312}{{\ttfamily arXiv:2110.04312
  [hep-ph]}}.

\bibitem{Mazde:2022sdx}
K.~Mazde and L.~Visinelli, ``{The Interplay between the Dark Matter Axion and
  Primordial Black Holes},'' \href{http://arxiv.org/abs/2209.14307}{{\ttfamily
  arXiv:2209.14307 [astro-ph.CO]}}.

\bibitem{Gregory:2013hja}
R.~Gregory, I.~G. Moss, and B.~Withers, ``{Black holes as bubble nucleation
  sites},'' \href{http://dx.doi.org/10.1007/JHEP03(2014)081}{{\em JHEP}
  {\bfseries 03} (2014) 081}, \href{http://arxiv.org/abs/1401.0017}{{\ttfamily
  arXiv:1401.0017 [hep-th]}}.

\bibitem{Burda:2015isa}
P.~Burda, R.~Gregory, and I.~Moss, ``{Gravity and the stability of the Higgs
  vacuum},'' \href{http://dx.doi.org/10.1103/PhysRevLett.115.071303}{{\em Phys.
  Rev. Lett.} {\bfseries 115} (2015) 071303},
  \href{http://arxiv.org/abs/1501.04937}{{\ttfamily arXiv:1501.04937
  [hep-th]}}.

\bibitem{Burda:2015yfa}
P.~Burda, R.~Gregory, and I.~Moss, ``{Vacuum metastability with black holes},''
  \href{http://dx.doi.org/10.1007/JHEP08(2015)114}{{\em JHEP} {\bfseries 08}
  (2015) 114}, \href{http://arxiv.org/abs/1503.07331}{{\ttfamily
  arXiv:1503.07331 [hep-th]}}.

\bibitem{Burda:2016mou}
P.~Burda, R.~Gregory, and I.~Moss, ``{The fate of the Higgs vacuum},''
  \href{http://dx.doi.org/10.1007/JHEP06(2016)025}{{\em JHEP} {\bfseries 06}
  (2016) 025}, \href{http://arxiv.org/abs/1601.02152}{{\ttfamily
  arXiv:1601.02152 [hep-th]}}.

\bibitem{Tetradis:2016vqb}
N.~Tetradis, ``{Black holes and Higgs stability},''
  \href{http://dx.doi.org/10.1088/1475-7516/2016/09/036}{{\em JCAP} {\bfseries
  09} (2016) 036}, \href{http://arxiv.org/abs/1606.04018}{{\ttfamily
  arXiv:1606.04018 [hep-ph]}}.

\bibitem{Canko:2017ebb}
D.~Canko, I.~Gialamas, G.~Jelic-Cizmek, A.~Riotto, and N.~Tetradis, ``{On the
  Catalysis of the Electroweak Vacuum Decay by Black Holes at High
  Temperature},'' \href{http://dx.doi.org/10.1140/epjc/s10052-018-5808-y}{{\em
  Eur. Phys. J. C} {\bfseries 78} no.~4, (2018) 328},
  \href{http://arxiv.org/abs/1706.01364}{{\ttfamily arXiv:1706.01364
  [hep-th]}}.

\bibitem{Gregory:2020cvy}
R.~Gregory, I.~G. Moss, and N.~Oshita, ``{Black Holes, Oscillating Instantons,
  and the Hawking-Moss transition},''
  \href{http://dx.doi.org/10.1007/JHEP07(2020)024}{{\em JHEP} {\bfseries 07}
  (2020) 024}, \href{http://arxiv.org/abs/2003.04927}{{\ttfamily
  arXiv:2003.04927 [hep-th]}}.

\bibitem{Hayashi:2020ocn}
T.~Hayashi, K.~Kamada, N.~Oshita, and J.~Yokoyama, ``{On catalyzed vacuum decay
  around a radiating black hole and the crisis of the electroweak vacuum},''
  \href{http://dx.doi.org/10.1007/JHEP08(2020)088}{{\em JHEP} {\bfseries 08}
  (2020) 088}, \href{http://arxiv.org/abs/2005.12808}{{\ttfamily
  arXiv:2005.12808 [hep-th]}}.

\bibitem{Gorbunov:2017fhq}
D.~Gorbunov, D.~Levkov, and A.~Panin, ``{Fatal youth of the Universe: black
  hole threat for the electroweak vacuum during preheating},''
  \href{http://dx.doi.org/10.1088/1475-7516/2017/10/016}{{\em JCAP} {\bfseries
  10} (2017) 016}, \href{http://arxiv.org/abs/1704.05399}{{\ttfamily
  arXiv:1704.05399 [astro-ph.CO]}}.

\bibitem{Mukaida:2017bgd}
K.~Mukaida and M.~Yamada, ``{False Vacuum Decay Catalyzed by Black Holes},''
  \href{http://dx.doi.org/10.1103/PhysRevD.96.103514}{{\em Phys. Rev. D}
  {\bfseries 96} no.~10, (2017) 103514},
  \href{http://arxiv.org/abs/1706.04523}{{\ttfamily arXiv:1706.04523
  [hep-th]}}.

\bibitem{Kohri:2017ybt}
K.~Kohri and H.~Matsui, ``{Electroweak Vacuum Collapse induced by Vacuum
  Fluctuations of the Higgs Field around Evaporating Black Holes},''
  \href{http://dx.doi.org/10.1103/PhysRevD.98.123509}{{\em Phys. Rev. D}
  {\bfseries 98} no.~12, (2018) 123509},
  \href{http://arxiv.org/abs/1708.02138}{{\ttfamily arXiv:1708.02138
  [hep-ph]}}.

\bibitem{Shkerin:2021zbf}
A.~Shkerin and S.~Sibiryakov, ``{Black hole induced false vacuum decay from
  first principles},'' \href{http://dx.doi.org/10.1007/JHEP11(2021)197}{{\em
  JHEP} {\bfseries 11} (2021) 197},
  \href{http://arxiv.org/abs/2105.09331}{{\ttfamily arXiv:2105.09331
  [hep-th]}}.

\bibitem{Shkerin:2021rhy}
A.~Shkerin and S.~Sibiryakov, ``{Black hole induced false vacuum decay: the
  role of greybody factors},''
  \href{http://dx.doi.org/10.1007/JHEP08(2022)161}{{\em JHEP} {\bfseries 08}
  (2022) 161}, \href{http://arxiv.org/abs/2111.08017}{{\ttfamily
  arXiv:2111.08017 [hep-th]}}.

\bibitem{Strumia:2022jil}
A.~Strumia, ``{Black holes don't source fast Higgs vacuum decay},''
  \href{http://arxiv.org/abs/2209.05504}{{\ttfamily arXiv:2209.05504
  [hep-ph]}}.

\bibitem{Inomata:2019ivs}
K.~Inomata, K.~Kohri, T.~Nakama, and T.~Terada, ``{Enhancement of Gravitational
  Waves Induced by Scalar Perturbations due to a Sudden Transition from an
  Early Matter Era to the Radiation Era},''
  \href{http://dx.doi.org/10.1103/PhysRevD.100.043532}{{\em Phys. Rev. D}
  {\bfseries 100} no.~4, (2019) 043532},
  \href{http://arxiv.org/abs/1904.12879}{{\ttfamily arXiv:1904.12879
  [astro-ph.CO]}}.

\bibitem{Inomata:2020lmk}
K.~Inomata, M.~Kawasaki, K.~Mukaida, T.~Terada, and T.~T. Yanagida,
  ``{Gravitational Wave Production right after a Primordial Black Hole
  Evaporation},'' \href{http://dx.doi.org/10.1103/PhysRevD.101.123533}{{\em
  Phys. Rev. D} {\bfseries 101} no.~12, (2020) 123533},
  \href{http://arxiv.org/abs/2003.10455}{{\ttfamily arXiv:2003.10455
  [astro-ph.CO]}}.

\bibitem{Das:2021wei}
S.~Das and A.~Hook, ``{Black hole production of monopoles in the early
  universe},'' \href{http://dx.doi.org/10.1007/JHEP12(2021)145}{{\em JHEP}
  {\bfseries 12} (2021) 145}, \href{http://arxiv.org/abs/2109.00039}{{\ttfamily
  arXiv:2109.00039 [hep-ph]}}.

\bibitem{Landau:1953um}
L.~D. Landau and I.~Pomeranchuk, ``{Limits of applicability of the theory of
  bremsstrahlung electrons and pair production at high-energies},'' {\em Dokl.
  Akad. Nauk Ser. Fiz.} {\bfseries 92} (1953) 535--536.

\bibitem{Migdal:1956tc}
A.~B. Migdal, ``{Bremsstrahlung and pair production in condensed media at
  high-energies},'' \href{http://dx.doi.org/10.1103/PhysRev.103.1811}{{\em
  Phys. Rev.} {\bfseries 103} (1956) 1811--1820}.

\bibitem{Gyulassy:1993hr}
M.~Gyulassy and X.-n. Wang, ``{Multiple collisions and induced gluon
  Bremsstrahlung in QCD},''
  \href{http://dx.doi.org/10.1016/0550-3213(94)90079-5}{{\em Nucl. Phys. B}
  {\bfseries 420} (1994) 583--614},
  \href{http://arxiv.org/abs/nucl-th/9306003}{{\ttfamily
  arXiv:nucl-th/9306003}}.

\bibitem{Arnold:2001ba}
P.~B. Arnold, G.~D. Moore, and L.~G. Yaffe, ``{Photon emission from
  ultrarelativistic plasmas},''
  \href{http://dx.doi.org/10.1088/1126-6708/2001/11/057}{{\em JHEP} {\bfseries
  11} (2001) 057}, \href{http://arxiv.org/abs/hep-ph/0109064}{{\ttfamily
  arXiv:hep-ph/0109064}}.

\bibitem{Arnold:2001ms}
P.~B. Arnold, G.~D. Moore, and L.~G. Yaffe, ``{Photon emission from quark gluon
  plasma: Complete leading order results},''
  \href{http://dx.doi.org/10.1088/1126-6708/2001/12/009}{{\em JHEP} {\bfseries
  12} (2001) 009}, \href{http://arxiv.org/abs/hep-ph/0111107}{{\ttfamily
  arXiv:hep-ph/0111107}}.

\bibitem{Arnold:2002ja}
P.~B. Arnold, G.~D. Moore, and L.~G. Yaffe, ``{Photon and gluon emission in
  relativistic plasmas},''
  \href{http://dx.doi.org/10.1088/1126-6708/2002/06/030}{{\em JHEP} {\bfseries
  06} (2002) 030}, \href{http://arxiv.org/abs/hep-ph/0204343}{{\ttfamily
  arXiv:hep-ph/0204343}}.

\bibitem{Besak:2010fb}
D.~Besak and D.~Bodeker, ``{Hard Thermal Loops for Soft or Collinear External
  Momenta},'' \href{http://dx.doi.org/10.1007/JHEP05(2010)007}{{\em JHEP}
  {\bfseries 05} (2010) 007}, \href{http://arxiv.org/abs/1002.0022}{{\ttfamily
  arXiv:1002.0022 [hep-ph]}}.

\bibitem{Kurkela:2014tla}
A.~Kurkela and U.~A. Wiedemann, ``{Picturing perturbative parton cascades in
  QCD matter},'' \href{http://dx.doi.org/10.1016/j.physletb.2014.11.054}{{\em
  Phys. Lett. B} {\bfseries 740} (2015) 172--178},
  \href{http://arxiv.org/abs/1407.0293}{{\ttfamily arXiv:1407.0293 [hep-ph]}}.

\bibitem{Harigaya:2013vwa}
K.~Harigaya and K.~Mukaida, ``{Thermalization after/during Reheating},''
  \href{http://dx.doi.org/10.1007/JHEP05(2014)006}{{\em JHEP} {\bfseries 05}
  (2014) 006}, \href{http://arxiv.org/abs/1312.3097}{{\ttfamily arXiv:1312.3097
  [hep-ph]}}.

\bibitem{Mukaida:2015ria}
K.~Mukaida and M.~Yamada, ``{Thermalization Process after Inflation and
  Effective Potential of Scalar Field},''
  \href{http://dx.doi.org/10.1088/1475-7516/2016/02/003}{{\em JCAP} {\bfseries
  02} (2016) 003}, \href{http://arxiv.org/abs/1506.07661}{{\ttfamily
  arXiv:1506.07661 [hep-ph]}}.

\bibitem{Kawasaki:2000en}
M.~Kawasaki, K.~Kohri, and N.~Sugiyama, ``{MeV scale reheating temperature and
  thermalization of neutrino background},''
  \href{http://dx.doi.org/10.1103/PhysRevD.62.023506}{{\em Phys. Rev. D}
  {\bfseries 62} (2000) 023506},
  \href{http://arxiv.org/abs/astro-ph/0002127}{{\ttfamily
  arXiv:astro-ph/0002127}}.

\bibitem{Hasegawa:2019jsa}
T.~Hasegawa, N.~Hiroshima, K.~Kohri, R.~S.~L. Hansen, T.~Tram, and
  S.~Hannestad, ``{MeV-scale reheating temperature and thermalization of
  oscillating neutrinos by radiative and hadronic decays of massive
  particles},'' \href{http://dx.doi.org/10.1088/1475-7516/2019/12/012}{{\em
  JCAP} {\bfseries 12} (2019) 012},
  \href{http://arxiv.org/abs/1908.10189}{{\ttfamily arXiv:1908.10189
  [hep-ph]}}.

\bibitem{Bodeker:2019ajh}
D.~B\"odeker and D.~Schr\"oder, ``{Equilibration of right-handed electrons},''
  \href{http://dx.doi.org/10.1088/1475-7516/2019/05/010}{{\em JCAP} {\bfseries
  05} (2019) 010}, \href{http://arxiv.org/abs/1902.07220}{{\ttfamily
  arXiv:1902.07220 [hep-ph]}}.

\bibitem{Mukaida:2022bbo}
K.~Mukaida and M.~Yamada, ``{Cascades of high-energy SM particles in the
  primordial thermal plasma},''
  \href{http://arxiv.org/abs/2208.11708}{{\ttfamily arXiv:2208.11708
  [hep-ph]}}.

\bibitem{Kurkela:2011ti}
A.~Kurkela and G.~D. Moore, ``{Thermalization in Weakly Coupled Nonabelian
  Plasmas},'' \href{http://dx.doi.org/10.1007/JHEP12(2011)044}{{\em JHEP}
  {\bfseries 12} (2011) 044}, \href{http://arxiv.org/abs/1107.5050}{{\ttfamily
  arXiv:1107.5050 [hep-ph]}}.

\bibitem{Harigaya:2014waa}
K.~Harigaya, M.~Kawasaki, K.~Mukaida, and M.~Yamada, ``{Dark Matter Production
  in Late Time Reheating},''
  \href{http://dx.doi.org/10.1103/PhysRevD.89.083532}{{\em Phys. Rev. D}
  {\bfseries 89} no.~8, (2014) 083532},
  \href{http://arxiv.org/abs/1402.2846}{{\ttfamily arXiv:1402.2846 [hep-ph]}}.

\bibitem{Bethe:1934za}
H.~Bethe and W.~Heitler, ``{On the Stopping of fast particles and on the
  creation of positive electrons},''
  \href{http://dx.doi.org/10.1098/rspa.1934.0140}{{\em Proc. Roy. Soc. Lond. A}
  {\bfseries 146} (1934) 83--112}.

\bibitem{Kim:2016iyf}
W.~Kim, ``{Origin of Hawking Radiation: Firewall or Atmosphere?},''
  \href{http://dx.doi.org/10.1007/s10714-016-2179-2}{{\em Gen. Rel. Grav.}
  {\bfseries 49} no.~2, (2017) 15},
  \href{http://arxiv.org/abs/1604.00465}{{\ttfamily arXiv:1604.00465
  [hep-th]}}.

\bibitem{Shi:1996ic}
X.-D. Shi, ``{Chaotic amplification of neutrino chemical potentials by neutrino
  oscillations in big bang nucleosynthesis},''
  \href{http://dx.doi.org/10.1103/PhysRevD.54.2753}{{\em Phys. Rev. D}
  {\bfseries 54} (1996) 2753--2760},
  \href{http://arxiv.org/abs/astro-ph/9602135}{{\ttfamily
  arXiv:astro-ph/9602135}}.

\bibitem{Shi:1999kg}
X.-D. Shi, G.~M. Fuller, and K.~Abazajian, ``{Neutrino mixing generated lepton
  asymmetry and the primordial He-4 abundance},''
  \href{http://dx.doi.org/10.1103/PhysRevD.60.063002}{{\em Phys. Rev. D}
  {\bfseries 60} (1999) 063002},
  \href{http://arxiv.org/abs/astro-ph/9905259}{{\ttfamily
  arXiv:astro-ph/9905259}}.

\bibitem{Dolgov:2002wy}
A.~D. Dolgov, ``{Neutrinos in cosmology},''
  \href{http://dx.doi.org/10.1016/S0370-1573(02)00139-4}{{\em Phys. Rept.}
  {\bfseries 370} (2002) 333--535},
  \href{http://arxiv.org/abs/hep-ph/0202122}{{\ttfamily arXiv:hep-ph/0202122}}.

\bibitem{Kawasaki:2002hq}
M.~Kawasaki, F.~Takahashi, and M.~Yamaguchi, ``{Large lepton asymmetry from Q
  balls},'' \href{http://dx.doi.org/10.1103/PhysRevD.66.043516}{{\em Phys. Rev.
  D} {\bfseries 66} (2002) 043516},
  \href{http://arxiv.org/abs/hep-ph/0205101}{{\ttfamily arXiv:hep-ph/0205101}}.

\bibitem{Kohri:1996ke}
K.~Kohri, M.~Kawasaki, and K.~Sato, ``{Big bang nucleosynthesis and lepton
  number asymmetry in the universe},''
  \href{http://dx.doi.org/10.1086/512793}{{\em Astrophys. J.} {\bfseries 490}
  (1997) 72--75}, \href{http://arxiv.org/abs/astro-ph/9612237}{{\ttfamily
  arXiv:astro-ph/9612237}}.

\bibitem{Matsumoto:2022tlr}
A.~Matsumoto {\em et~al.}, ``{EMPRESS. VIII. A New Determination of Primordial
  He Abundance with Extremely Metal-Poor Galaxies: A Suggestion of the Lepton
  Asymmetry and Implications for the Hubble Tension},''
  \href{http://arxiv.org/abs/2203.09617}{{\ttfamily arXiv:2203.09617
  [astro-ph.CO]}}.

\bibitem{DOnofrio:2014rug}
M.~D'Onofrio, K.~Rummukainen, and A.~Tranberg, ``{Sphaleron Rate in the Minimal
  Standard Model},''
  \href{http://dx.doi.org/10.1103/PhysRevLett.113.141602}{{\em Phys. Rev.
  Lett.} {\bfseries 113} no.~14, (2014) 141602},
  \href{http://arxiv.org/abs/1404.3565}{{\ttfamily arXiv:1404.3565 [hep-ph]}}.

\bibitem{Ai:2022kqm}
W.-Y. Ai, J.~S. Cruz, B.~Garbrecht, and C.~Tamarit, ``{Instability of bubble
  expansion at zero temperature},''
  \href{http://arxiv.org/abs/2209.00639}{{\ttfamily arXiv:2209.00639
  [hep-th]}}.

\bibitem{Harigaya:2019tzu}
K.~Harigaya, K.~Mukaida, and M.~Yamada, ``{Dark Matter Production during the
  Thermalization Era},'' \href{http://dx.doi.org/10.1007/JHEP07(2019)059}{{\em
  JHEP} {\bfseries 07} (2019) 059},
  \href{http://arxiv.org/abs/1901.11027}{{\ttfamily arXiv:1901.11027
  [hep-ph]}}.

\bibitem{Drees:2021lbm}
M.~Drees and B.~Najjari, ``{Energy spectrum of thermalizing high energy decay
  products in the early universe},''
  \href{http://dx.doi.org/10.1088/1475-7516/2021/10/009}{{\em JCAP} {\bfseries
  10} (2021) 009}, \href{http://arxiv.org/abs/2105.01935}{{\ttfamily
  arXiv:2105.01935 [hep-ph]}}.

\bibitem{Drees:2022vvn}
M.~Drees and B.~Najjari, ``{Multi-Species Thermalization Cascade of Energetic
  Particles in the Early Universe},''
  \href{http://arxiv.org/abs/2205.07741}{{\ttfamily arXiv:2205.07741
  [hep-ph]}}.

\end{thebibliography}\endgroup

\end{document}